\documentclass[aps,prb,twocolumn,amsmath,amssymb,superscriptaddress,floatfix]{revtex4-2}

\usepackage{amsmath}
\usepackage{xcolor}
\usepackage{graphicx}
\usepackage{dcolumn}
\usepackage{bm}
\usepackage{verse}

\newcommand{\CNN}{Universit{\'e} Paris-Saclay, CNRS, Centre de Nanosciences et de Nanotechnologies, 91120 Palaiseau, France}
\newcommand{\IMEC}{IMEC, Kapeldreef 75, B-3001 Leuven, Belgium}

\begin{document}


\title {Spin-torque induced wall motion in perpendicularly magnetized discs: ballistic versus oscillatory behavior}

\author{Paul Bouquin}
 \email{paul.bouquin@u-psud.fr}
\affiliation{\CNN}
\affiliation{\IMEC}
\author{Joo-Von Kim}
\affiliation{\CNN}
\author{Olivier Bultynck}
\author{Siddharth Rao}
\author{Sebastien Couet}
\author{Gouri Sankar Kar}
\affiliation{\IMEC}
\author{Thibaut Devolder}
\email{thibaut.devolder@u-psud.fr}
\affiliation{\CNN}

\date{\today}

\begin{abstract}
We use time-resolved measurement and modeling to study the spin-torque induced motion of a domain wall in perpendicular anisotropy magnets. In disc of diameters between 70 and 100 nm, the wall drifts across the disc with pronounced back-and-forth oscillations that arise because the wall moves in the Walker regime. Several switching paths occur stochastically and lead to distinct switching durations. The wall can cross the disc center either in a ballistic manner or with variably marked oscillations before and after the crossing. The crossing of the center can even occur multiple times if a vertical Bloch line nucleates within the wall. The wall motion is analyzed using a collective coordinate model parametrized by the wall position $q$ and the tilt $\phi$ of its in-plane magnetization projection. The dynamics results from the stretch field, which describes the affinity of the wall to reduce its length and the wall stiffness field describing the wall tendency to reduce dipolar energy by rotating its tilt. The wall oscillations result from the continuous exchange of energy between to the two degrees of freedom $q$ and $\phi$. 
The stochasticity of the wall dynamics can be understood from the concept of the retention pond: a region in the $q-\phi$ space in which walls are transiently bound to the disc center. Walls having trajectories close to the pond must circumvent it and therefore have longer propagation times. The retention pond disappears for a disc diameter of typically 40 nm: the wall then moves in a ballistic manner irrespective of the dynamics of its tilt. The propagation time is then robust against fluctuations hence reproducible.
\end{abstract}

\maketitle

\section{Introduction}
The understanding of magnetization reversal in nanostructures is a vast and long-standing field of research that is of considerable application interest. The magnetization reversal can be induced by a plethora of stimuli, from the classical magnetic fields to the various torques that emerge from the coupling of the magnetization with the crystal lattice \cite{novosad_novel_2000}, with the electric field \cite{ohno_electric-field_2000}, with light \cite{beaurepaire_ultrafast_1996} or with the electric current \cite{chappert_emergence_2007}. Among these torques, the spin-transfer-torque (STT) is particularly relevant for ultrathin magnets with perpendicular magnetic anisotropy (PMA).
The manner in which STT switches the magnetization depends largely on the system size. When the lateral size of an ultrathin magnet is comparable or smaller that the domain wall width the reversal happens in a quasi-coherent manner \cite{khvalkovskiy_basic_2013, bouquin_size_2018}. At larger sizes, the reversal implies the nucleation and the subsequent propagation of domain walls \cite{you_switching_2014, chaves-oflynn_thermal_2015}. The switching dynamics in the coherent regime is well understood \cite{sun_spin-current_2000} even in the presence of thermal fluctuations \cite{butler_switching_2012, pinna_spin-transfer_2013, tomita_unified_2013} but it is seldom observed in practical systems \cite{bernstein_nonuniform_2011, sun_spin-torque_2013, hahn_time-resolved_2016}. The domain-wall-based switching regime is comparatively more complex as it implies at least a nucleation event which is influenced by the geometry, the materials microstructure, the detail of its properties \cite{pizzini_chirality-induced_2014} and the thermal fluctuations. The bottleneck character of the nucleation step \cite{munira_calculation_2015, visscher_instability_2016} has prevented an exhaustive study of the subsequent wall propagation. In elongated nanostructures, the STT-induced wall motion was found \cite{devolder_time-resolved_2016} to happen in the precessional Walker regime with back-and-forth oscillation. In circular discs, single-shot time-resolved electrical measurements evidenced that the propagation time scales inversely with the spin-torque amplitude \cite{hahn_time-resolved_2016, tomita_high-speed_2011} and approximately linearly with the disc diameter \cite{devolder_material_2018} with a strong event-to-event variability even in the deep sub-100~nm size regime \cite{sun_spin-torque_2013, hahn_time-resolved_2016, devolder_size_2016}. The understanding of the speed of magnetization reversal and the variability thereof is still incomplete in PMA nanostructures. 

In this paper, we study part of this problem: we let aside the nucleation phenomenon and focus on the subsequent dynamics of a domain wall placed in a circular magnet and then submitted to STT. We expand a recent study \cite{bouquin_stochastic_2021} where we evidenced that stochastic processes in wall motion were inducing counterintuitive temporal pinning of the wall near the disc center with a strong oscillatory character. These features were also reported independently in a modeling paper \cite{statuto_micromagnetic_2021}. The present paper deepens our understanding of the domain wall motion by first reporting more extensive time-resolved single-shot electrical measurements, and then by further developing the analytical models and benchmarking them with micromagnetics. The model is used to design strategies applicable to reduce the variability of the domain wall propagation dynamics.

The paper is organized as follows. We start with the experimental results (§\ref{exp}). Section~\ref{sectionmumag} then describes the domain wall dynamics when computed using micromagnetics. Section \ref{CollectiveCoordModel} is devoted to a collective coordinate model in which the dynamics is simplified and aggregated in only two degrees of freedom: the position $q$ of the wall and the tilt $\phi$ of the in-plane component of the magnetization within the wall. The physical meaning of the model is described in Section \ref{ApproximateAnalyticalSection}. It is compared to micromagnetics in section~\ref{SectionComparaisonDesModeles} and discussed in section~\ref{sectiondiscussion}, before concluding.

\section{Electrical signatures of switching}
\label{exp}
\subsection{Materials and experimental methods} 

In this first section, our objective is to identify the key features of the dynamics of a domain wall responding to STT. We use time-resolved conductance measurements in a magnetic tunnel junction (MTJ). We have selected supersoft samples, i.e. optimized for very easy domain wall propagation. The primary requirement on the free layer (FL) is to avoid non-uniformities that would perturb the wall dynamics. The structural non-uniformities --inducing pinning sites-- are best avoided when the FL growth can be optimized with a total freedom on the material buffer. This is possible only for \textit{top-pinned} MTJs. The non-uniformities can also arise from the (unavoidably non-uniform \cite{devolder_offset_2019}) stray field that emanate from the reference layers of the MTJ. To minimize the influence of the stray field, we work with FL possessing a low moment as well as large thickness, which ends by a cap of 8\r{A} of FeCoB in contact with the MgO tunnel oxide. These requirements prevent the use of the state-of-the-art material systems optimized for memory applications ; our MTJs exhibit  modest transport properties, with a magneto-resistance of 80\% for a resistance-area product of $RA=9.6 ~\Omega.\mu\textrm m^2$. Besides, the stability of the reference layer is somewhat insufficient in the presence of current, such that we can only study the high resistance to low resistance switching transition. 

The MTJs are patterned into elongated as well as disc-shaped devices. The first devices are rectangles of size $80\times280~\textrm{nm}^2$; the corners are rounded with a radius of curvature of 20 nm: these samples illustrate the model case of the wall propagation in a long stripe. The discs have diameters in the 70-100 nm interval, i.e. a region in which domain-wall based reversal is expected \cite{bouquin_size_2018}. The quoted diameters are given by the ratio of the $RA$ by the device resistance $R$: it is the area in which the FL is metallic. The magnetic diameter in which the FL has good magnetic properties might be smaller.

\begin{figure}[t!]
\begin{center}
\includegraphics[width=9.2 cm]{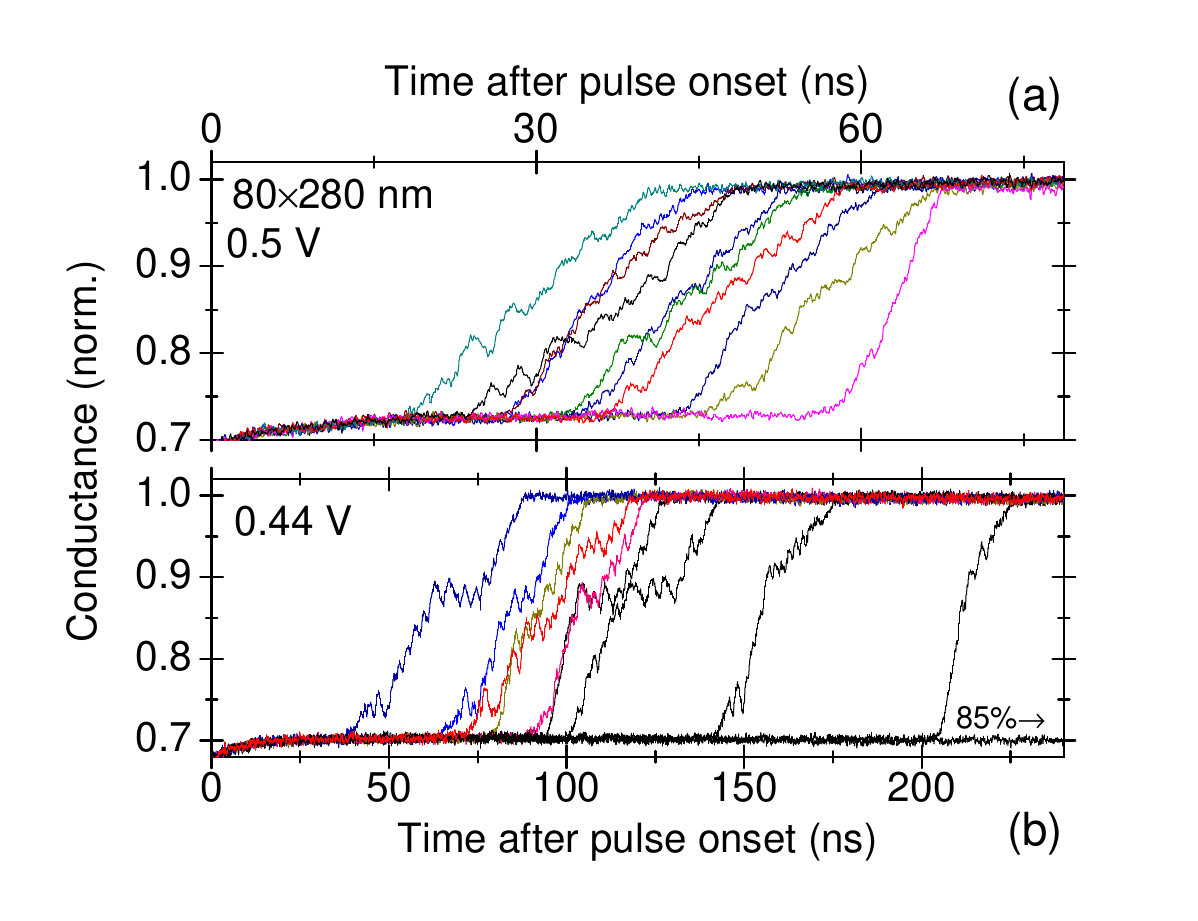}
\caption{Conductance signatures of various spin-torque induced switching events as a result of voltage pulses of (a): 500 mV and (b): 440 mV all measured on the same rectangular device of size of $\approx80\times280~\textrm{nm}^2$ in ambiant temperature conditions. In panel (b) the applied voltage is close to the quasi-static switching threshold and 85\% of the voltage steps did not yet induce a switching after 250 ns.}
\label{wavy80nm}
\end{center}
\end{figure}

The measurement set-up aims at applying fast rising voltage steps with maximally flat voltage plateaus. It monitors the time-resolved device conductance under constant voltage with a bandwidth of 15 GHz. The switching in rectangular devices and in disc-shaped devices are illustrated in Fig.~\ref{wavy80nm} and ~\ref{wavy100nm}. In the first 5 ns after the pulse onset, the conductance of the antiparallel states (initial states) rises asymptotically in a reproducible manner. This arises mainly from the imperfect flatness of the applied voltage plateau and from the slight voltage dependence of the conductance. This first 5 ns period is also the one during which the MTJs heats up by Joule effect 
; We will assume that the subsequent evolution of the conductance reflects purely the magnetic moment $\langle m_z \rangle$ of the free layer in a proportional manner, at \textit{constant} voltage and temperature.

\subsection{Dynamics in rectangular elongated devices} 
The switching in the rectangular devices is reported in Fig.~\ref{wavy80nm}. At 500 mV, i.e.~a couple of dBs above the quasi-static switching threshold (400 mV), the switching proceeds in the usual \cite{devolder_single-shot_2008, hahn_time-resolved_2016, bultynck_instant-spin_2018} two steps [Fig.~\ref{wavy80nm}(a)]. The magnetization seems first quiet during a stochastically varying incubation delay. Then the conductance grows in ramp-like manner, on which strong oscillations are superimposed most of the time, until it saturates to the conductance of the final state. Note despite their clear event-to-event variability, the conductance oscillations are real, i.e. way above the noise level of the experiment. This oscillatory behavior was formerly interpreted \cite{devolder_time-resolved_2016, devolder_material_2018} as the signature that the wall propagation was occurring in the regime above the Walker breakdown, i.e. the domain wall advances in average but through a perpetual back and forth cyclic motion. We will confirm this assertion in the next sections. 

When the voltage is reduced to near the quasi-static threshold [Fig.~\ref{wavy80nm}(b)], the switching probability and its rate both decrease. The conductance oscillations are still present and still exhibit a large variability. However the system shows now a propensity to oscillate a longer time at intermediate conductance levels distributed in the region where $-0.2 \leq \langle m_z \rangle \leq 0.55$, where $m_z$ is the out-of-plane component of the magnetization. We emphasize that this oscillatory character, as well as the propensity to oscillate a longer time at in the midway region, are both \textit{very} sensitive to the out-of-plane field $H_z$ that is applied to the rectangular devices. The switching curves become featureless ramps (not shown) if a field of 15 mT is added is either direction away from the center of the $R(H_z)$ minor loop.  Previous time-resolved studies may have missed to observe this oscillatory character because of its extreme sensitivity to the field.

\subsection{Dynamics in disc-shaped devices} 
\begin{figure*}[t!]
\begin{center}
\includegraphics[width=18 cm]{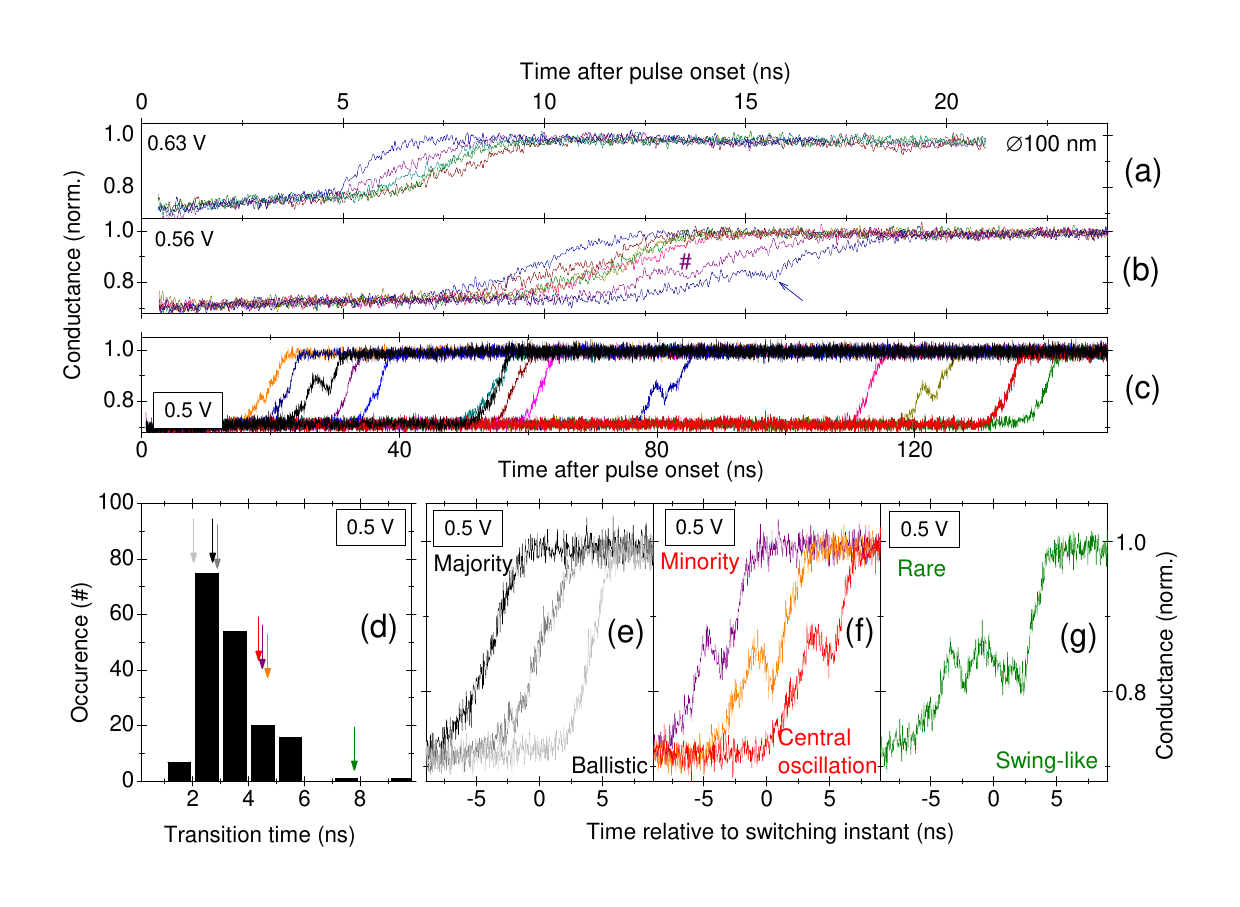}
\caption{Conductance signatures of various spin-torque induced switching events as a result of voltage pulses of (a): 630 mV, (b): 560 mV and (c-g): 500 mV on a circular device of electrical diameter of $\approx$100 nm. In (b) the arrow (respectively the hashtag) points at a switching event in which the reversal rate seems to slow down (respectively oscillate) at midway. At the lowest applied voltage, three different signatures are observed : drift-like one-way crossing of the midway conductance with featureless ramp-like shape [gray, (e)],  crossing of the midway with one single pronounced oscillation [red, (f)] and multiple crossing of the midway conductivity [green, (g)]. In the panels (e, f), the side curves were horizontally offset by $\pm4~\textrm{ns}$ to avoid overlaps. The different classes of switching events have transition durations whose histogram is depicted in panel (d). }
\label{wavy100nm}
\end{center}
\end{figure*}

We now move to circular devices. We focus on a representative device of diameter 100 nm, but a very similar behavior is found in the investigated interval of sizes (70-100 nm). At large voltage [630 mV, i.e. 1.4 times the quasi-static switching threshold of 440 mV, Fig.~\ref{wavy100nm}(a)], the magnetization switches after an average incubation delay $\langle t_0 \rangle$ of 5.6 ns, within an average transition time $\langle \tau \rangle$ of 2.5 ns. At this large voltage, the conductance waveforms are featureless ramps that can be reasonably well fitted by $\textrm{erf} [(t-t_0) / \tau]$ functions to extract $t_0$ and $\tau$.

When reducing the voltage, both the incubation delay and the transition time increase on average. 
A careful examination of the switching curves indicates that their variability is enhanced when the conductance is near the conductance midway point (i.e. near  $\langle m_z \rangle \approx 0$): some slowing down of the rate can sometimes be observed [see the arrow in Fig.~\ref{wavy100nm}(b)]. An oscillation can also occasionally be observed [see the purple hashtag]. Finally, an apparent acceleration is sometimes evidenced [see the two green curves that overtake other curves within Fig.~\ref{wavy100nm}(b)]. At the two highest voltages, we never perceived more than one clear oscillation in the time-resolved curves. There might either never be more than one oscillation, or most probably the tiniest oscillations are hidden by the instrumental noise, which is comparatively larger because less current is flowing in circular devices than in the elongated devices that had a thrice larger surface.

Finally, the time-resolved conductance curves change \textit{qualitatively} when the voltage is reduced to 10\% above the quasi-static switching threshold Fig.~\ref{wavy100nm}(c)]: at 500 mV, one can clearly identify 3 categories of switching events with differing shapes and transition times. There is no correlation between the incubation delays and the transition times. (i) The most occurring events correspond to featureless ramps of the conductance waveforms, with transition times between typically 1 and 3 ns [Fig.~\ref{wavy100nm}(e)] ; they shall be referred as the "ballistic events". (ii) For a substantial minority of events (20\% probability), a single strong oscillation is observed when the conductance is at midway between the initial and final states. The conductance first passes above the midway value, then it reduces for $1.3\pm0.1~\textrm{ns}$ until it finally rises again till saturation, such that the transition time is now typically between 4 and 6 ns. These events shall be referred as the "central oscillation" events.  (iii) Finally, in rare occasions ($\approx$2\% probability) the conductance crosses the midway value multiple times [Fig.~\ref{wavy100nm}(g)] and the transition time can then exceed 7 ns. These events shall be referred as the "multiple swing" events. They skew the distribution of transition times to the higher values [Fig.~\ref{wavy100nm}(d)].

Before starting the theoretical part of this article, let us summarize our experimental findings. In elongated devices, the conductance waveforms exhibit a clear oscillatory character that suggests a reversal using a domain wall that sweeps through the sample in a regime above the Walker breakdown. 
In disc-shaped devices submitted to high voltages, the conductance waveforms are monotonic with fluctuations of both the incubation delay and the transition time. When reducing the voltage, the waveforms gain an increasing complexity, particularly when the conductance is at midway between the initial and final states. This complexity culminates when the applied voltage is slightly above the switching threshold, with three identified behaviors. In most switching events, the conductances evolves monotonically, as if the domain wall was sweeping in a regular manner through the device. In 20\% of the cases, the conductance exhibits a pronounced oscillation at midway.  We will see that this happens when the wall performs one oscillation about the disc center. In rare occasions, the conductance oscillates more than once in the midway value; We will see that this happens when a vertical Bloch line \cite{krizakova_study_2019} nucleates within the wall.

\section{STT-induced Domain wall motion within a disc in the micromagnetism theory}
\label{sectionmumag}
\subsection{Geometry and material parameters}
We consider a thin disc with perpendicular magnetic anisotropy. We shall study the domain wall dynamics within this disc at zero temperature in the absence of applied field, as resulting from a Slonczewski-like STT induced by a current carrying a spin polarization $P$ directed along the magnetic easy axis $(z)$ of the disc. This situation is meant to mimic the evolution of a DW placed in a CoFeB/MgO/CoFeB-based MTJ that is biased by a constant and uniform voltage, as in experiments. The Landau-Lifshitz-Gilbert-Slonczewski (LLGS) equation is solved using the micromagnetic solver mumax$^3$ \cite{vansteenkiste_design_2014}. We assume: a film thickness $d=2 \ \mathrm{nm}$, a magnetization $M_s=1.2 \ \mathrm{MA}/\mathrm{m}$, a magneto-crystalline anisotropy field $H_k=1.566 \ \mathrm{MA}/\mathrm{m}$, a damping parameter $\alpha=0.01$ and an exchange stiffness of $A_\mathrm{ex}=20 \ \mathrm{pJ}/ \mathrm{m}$. 

The STT is implemented using the torque-to-voltage correspondence described in ref.~\cite{bouquin_size_2018}. We will see that there is no obvious characteristic voltage to be defined in the dynamics of wall motion. However, it will be insightful to compare the strength of our applied STT to the voltage $V_\mathrm{c}$: 
\begin{equation}
V_{\mathrm{c}}= \frac{2 \alpha e \mathcal{A}R_\perp d \mu_0 M_s H_\mathrm{k,eff}^\mathrm{disc}}{P \hbar}~,
\label{macrospinVc}
\end{equation}
which is the voltage that would be needed to destabilize a uniformly magnetized state at zero temperature~\cite{bouquin_size_2018}; $V_{\mathrm{c}}$ is thus closely related to the voltage needed to initiate the magnetization reversal and to potentially nucleate a domain wall within the disc~\cite{you_switching_2014}. In Eq.~\ref{macrospinVc} $\mathcal{A}R_\perp$ is the resistance-area product of the MTJ when the magnetization is perpendicular to the spin-polarization axis and $H_\mathrm{k,eff}^\mathrm{disc}$ is the effective anisotropy field of the disc.

In the theory parts of this paper, we focus on two specific disc diameters: $2R=40 \ \mathrm{nm}$ or $2R=80 \ \mathrm{nm}$ that appear to be well suited to reveal the wealth of the wall dynamics. The largest size will be qualitatively compared to the experimental data. In micromagnetics, the complexity of the wall dynamics grows substantially for $2R > 80 \ \mathrm{nm}$ \cite{bouquin_size_2018} and will not be reported in detail here. For instance the magnetization within the wall gets substantially non-uniform for $2R = 150 \ \mathrm{nm}$ while for $2R \geq 200 \ \mathrm{nm}$ a nucleation event occurs at a position in the still unreversed domain before the wall reaches that same position. Except in the supplementary material, we study the domain wall dynamics at applied voltages equal to the respective instability thresholds. 

\subsection{Initialization of the magnetization states: straight or curved domain walls} 
We study the dynamics of a wall placed within the disc in an ad-hoc manner. This dynamics depends on the initial state of the wall. Note that our experiments were performed at room temperature; unfortunately in the experiments, we had strictly no handle on the wall state when nucleated. To extract now the main essence of the domain wall dynamics, we now focus on two numerical ways of preparing the wall: an initially \textit{straight} wall and an initially \textit{curved} wall whose lower energy \cite{chaves-oflynn_thermal_2015} leads to a simpler dynamics. 

We first consider the dynamics of an initially straight wall having a magnetization profile that matches with the static case \cite{thiaville_domain_2002}, i.e. with a width $\pi \Delta$ where $\Delta=\sqrt{A_\mathrm{ex} / K_\mathrm{eff}}$ with $K_\mathrm{eff}$ the effective anisotropy energy of the film. The in-plane projection of the magnetization is set to have a uniform tilt $\phi (x,y)=\phi_0 $. 
The supplementary material shows that when initialized with a straight shape, the wall bents immediately. The bending breathes substantially at the beginning of the simulation. An adequate procedure (see suppl. material) lets the bending relax in order to prepare an \textit{optimally bent} wall to be used as a starting micromagnetic state (first insets in the panels (a) of Fig.~\ref{LLGS80nm} and \ref{LLGS40nm}]). This suppresses empirically the transient breathing dynamics of the wall while preserving the other dynamical features. The breathing dynamics seems independent from the other degrees of freedom of the wall, and will therefore be commented on only in the supplementary material.
In the optimally bent wall, the magnetization tilt $\phi$ along the wall length is no longer strictly uniform. In this case we simply define the wall tilt $\phi$ as the tilt at center of the wall length. We also redefine the wall 'position' $q$, by assimilating it to the position $q$ of a fictitiously straight wall that would yield the same total magnetic moment $\langle m_z \rangle$ as that of the curved wall.

\subsection{Domain wall dynamics within a disc of diameter of 80 nm} 
Figure~\ref{LLGS80nm}(a) illustrates the motion of a  wall with an optimal initial curvature and an initial tilt $\phi_0 = 30~\textrm{deg.}$ within a disc of diameter 80 nm at the applied voltage $V_{\mathrm{c}}$. Animations can be found in the supplementary material. The wall sweeps through the disc, acquires a straight shape when at the disc center and then bents again. The magnetic moment $\langle m_z \rangle$ reverses in a few ns [Fig.~\ref{LLGS80nm}(b)]. Several points are worth noticing in addition to this global drift of the DW.
\subsubsection{Overall motion: domain wall drift and superimposed oscillations}
The first important point is that the wall velocity and the magnetization tilt $\phi(t)$ [i.e. the color in Fig.~\ref{LLGS80nm}(a)] are coupled: $\phi(t)$ increases monotonously during the DW motion towards the disc center and this modulates the wall velocity. The wall switches periodically from the Bloch state ($\phi= \frac{\pi}{2}~[\pi]$) to the N\'eel state ($\phi= \pi~[\pi]$). For 80 nm diameter discs, the DW advances (i.e. $\dot q > 0$) when near the N\'eel state and moves backward when near the Bloch state [see the lower snapshots in Fig.~\ref{LLGS80nm}(b)], at the noticeable exception of when the wall crosses the center of the disc.
This back-and-forth motion of the DW is superimposed on the global drift of the DW that sweeps through the disc. The oscillation has a more pronounced amplitude when the wall approaches the disc center (when $\langle m_z \rangle \approx 0$). As this same position the oscillation of the moment and the rate of change $\dot \phi(t)$ of the tilt slow down (see the videos in the supplementary material). 

\subsubsection{Sensitivity to the initial conditions} 
\begin{figure*}[t!]
\begin{center}
\includegraphics[width=14 cm]{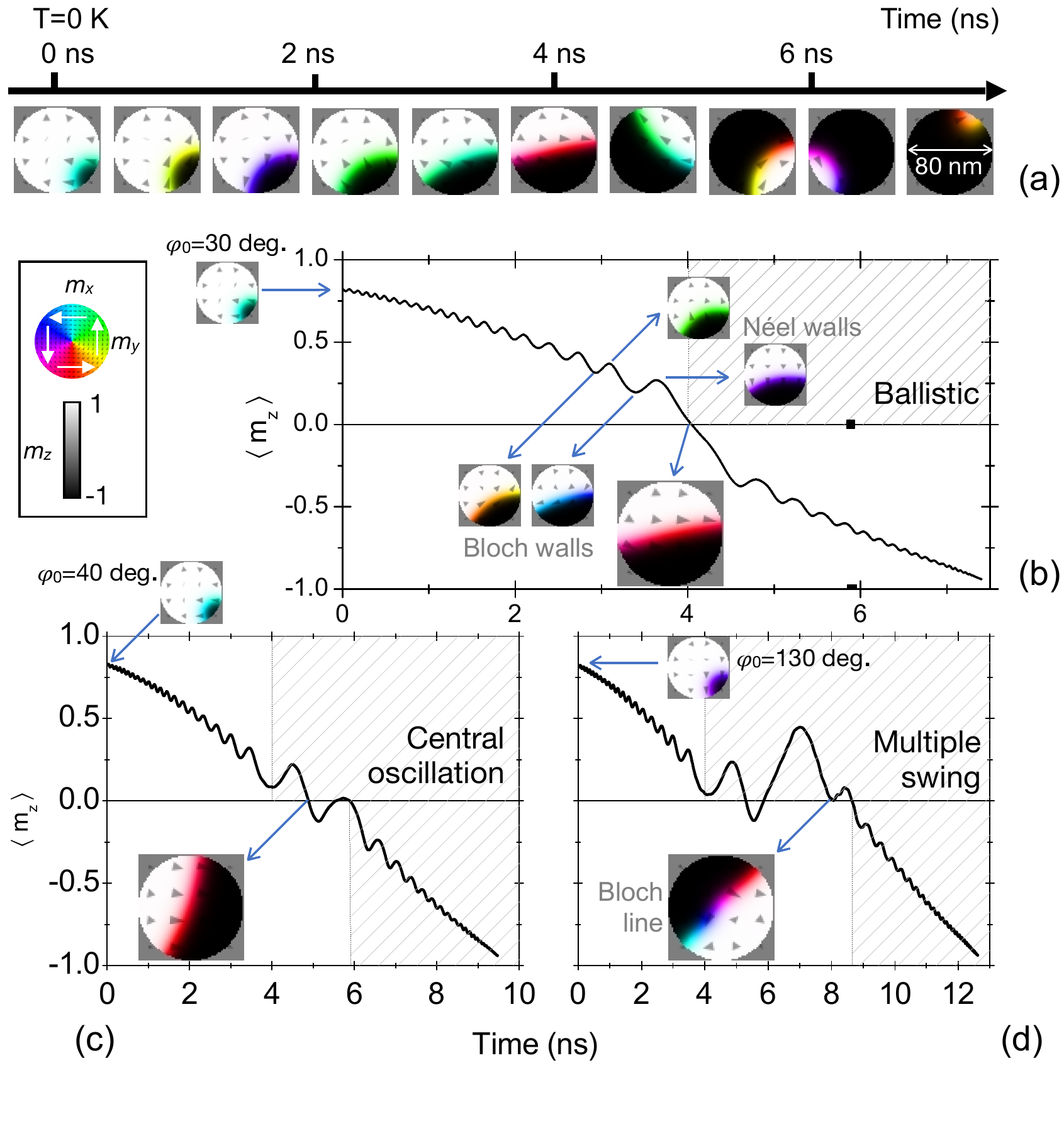}
\caption{Micromagnetic simulations of the DW dynamics within a 80-nm disc for walls with optimal initial curvatures. (a) Snapshots of the initial magnetization state with $\{\phi_0,~ q_0\} = \{30~\textrm{deg.},~10~\textrm{nm}\}$ and its subsequent evolution under a voltage equal to $V_c$ 
. (b) Resulting evolution of the spatial average of the $m_z$ component of the magnetization. The micromagnetic snapshots are taken at the extrema of $\langle m_z \rangle$. They illustrate the N\'eel (respectively Bloch) character of the wall when the DW velocity changes from negative to positive (resp. positive to negative). The panels (c) and (d) present similar plots calculated for different initial states in which in the initial tilts were rotated by respectively $\Delta \phi_0=+10~\textrm{deg.}$ and then by $\Delta  \phi_0=+90~\textrm{deg}$. The dashed areas help to better reveal the resulting changes in switching duration. Inset in panel (d): the non-uniformity of the magnetization tilt within the wall (seen by a multicolored wall with the label "Bloch line") correlates with the occurrence of a switching scenario in which the DW performs multiple crossing of the disc center in a swing-like manner. }
\label{LLGS80nm}
\end{center}
\end{figure*}

The second noticeable point is the large sensitivity of the dynamics to the initial conditions [Fig.~\ref{LLGS80nm}(b,c and d)]: minute changes of either the chosen initial DW position ($q_0$, not shown), of the initial tilt angle ($\phi_0$) alter substantially the dynamics when the wall later arrives in the center of the disc. We believe that this strong sensitivity to initial conditions contributes to the fluctuations of the transition time observed in room temperature experiments.

We have identified three scenarios in these micromagnetic simulations: a drift-like one-way "ballistic" crossing, a one-way crossing with pre- and post-crossing pauses and a swing-like crossing with multiple attempts. Most initial conditions $\{q_0, \phi_0\}$ lead to a drift-like one-way ballistic crossing: the wall sweeps through the center without stopping there. The time-resolved magnetic moment is nearly linear [Fig.~\ref{LLGS80nm}(b)]. For some specific other initial conditions, the wall stops on either sides of the disc diameter before and after crossing the center [Fig.~\ref{LLGS80nm}(c)], providing the impression of a central oscillation. In both scenarios the DW crosses the center while being essentially in a Bloch-type configuration. Note that this holds for a disc diameter of 80 nm but shall no longer hold for the smaller disc diameter. 
Occasionally, the DW swings several times back-and-forth in the vicinity of the disc diameter and crosses the center multiple times before finally leaving the center [Fig.~\ref{LLGS80nm}(d)]. In this last scenario, the tilt $\phi$ of the magnetization within the wall gets substantially non-uniform when the wall is in the vicinity of the disc center: a partial Bloch line is created within the wall [inset in Fig.~\ref{LLGS80nm}(d)]. These three micromagnetic scenarios recall to the three categories of electrical signatures of the transition observed experimentally voltages just above the switching threshold in the 70-100 nm samples [Fig.~\ref{wavy100nm}(e-g)].

\subsubsection{Minor features of the domain wall dynamics}
In addition to its gradual drift, its fast position oscillation and the pronounced or absent breathing of its curvature, the DW also rotates around the center of the disc [Fig.~\ref{LLGS80nm}(a)]. The gyration speed is minimal when $\phi$ is quasi-uniform; the gyration speed seems to increase coincidently with the non-uniformity of the tilt. This gyration of the wall recalls the gyration of non-trivial magnetic textures (vortices, skyrmions,...). Unfortunately the details of the gyration of the wall depend on how the (ideally perfectly circular) disc is deformed by mapping it on the square simulation grid of the micromagnetic solver. Staircase artefacts prevent an objective analysis of the gyration dynamics.

\subsection{Domain wall dynamics within a disc of diameter of 40 nm}
\begin{figure}
\includegraphics[width=8 cm]{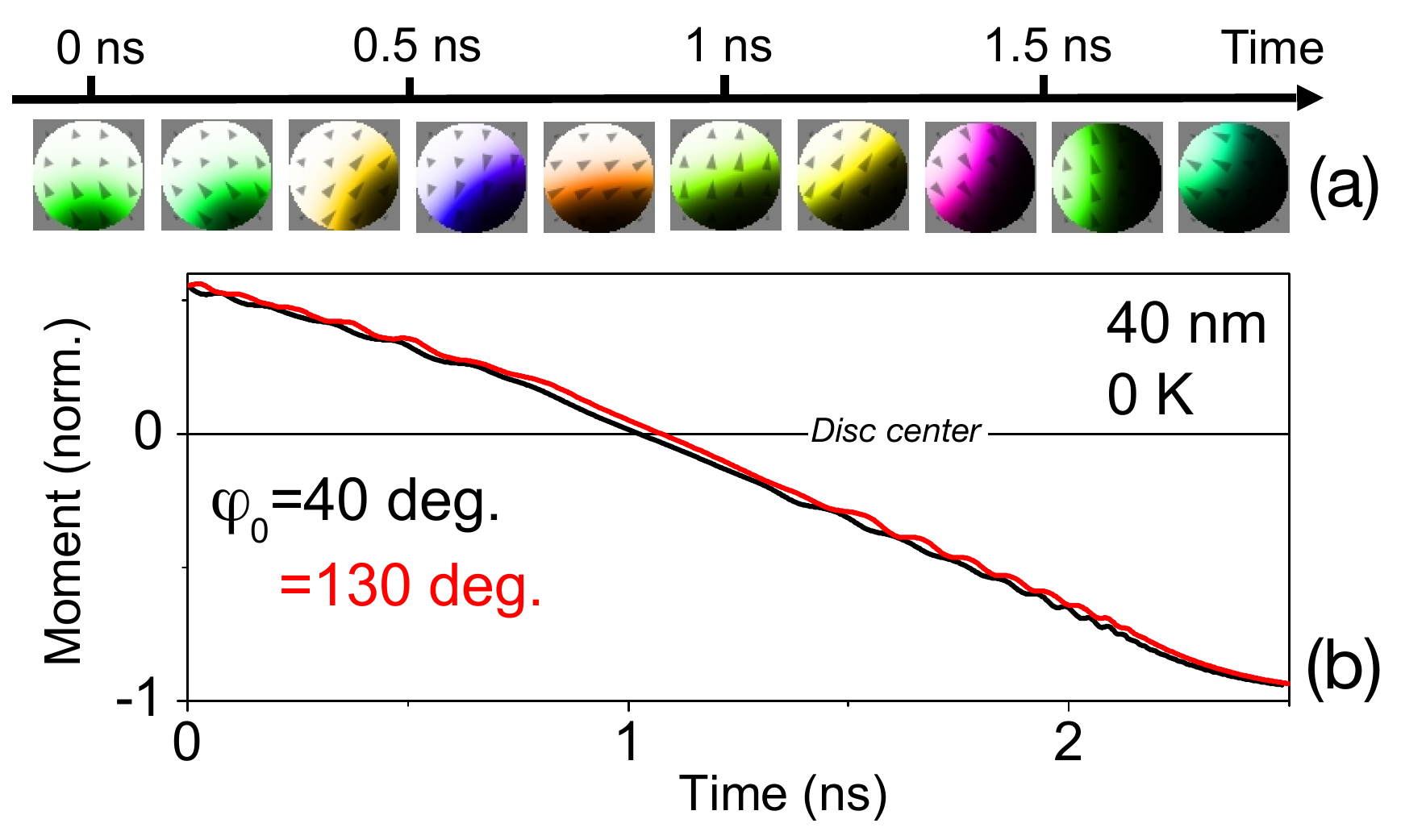}
\caption{Micromagnetic simulations of the DW dynamics within a 40-nm disc for a wall with optimal initial curvature. a) Snapshots of an initial magnetization state with $\{\phi_0, q_0\} = \{40~\textrm{deg.}, 10~\textrm{nm}\}$ and its subsequent evolution under an applied voltage equal to $V_c$ 
. b) Resulting evolution of the spatial average of the $m_z$ component of the magnetization for the same simulation and for one with an initial tilt angle incremented by 90 degrees.}
\label{LLGS40nm}
\end{figure}

Let us now consider smaller discs. Figure~\ref{LLGS40nm} gathers the main features of the DW dynamics within a smaller disc of diameter 40 nm when the wall is initialized with an optimal curvature. The main qualitative features of the dynamics formerly observed in the larger disc are preserved: there is still a gradual drift of the DW on which a fast position oscillation is superimposed. 
However the reduction of the diameter induces quantitative differences on the dynamics. In the 40 nm disc, the oscillatory character of the domain wall velocity is much less pronounced and can hardly be noticed [Fig.~\ref{LLGS40nm}(b)], especially near the disc center where the curves look very linear. Among the 3 scenarios previously identified on the 80 nm disc, only the drift-like one-way "ballistic" crossing is still observed when the diameter is 40 nm. Another difference is that the initialization conditions $\{\phi_0, q_0\}$ of the DW do no longer have a strong impact on the dynamics of the total moment. In 40 nm discs, the walls cross the disc center in a manner that does not depend much on their N\'eel of Bloch character.

So far we have described the DW dynamics by solving numerically the LLGS equations and by comparing the switching scenarios to the electrical signatures of time-resolved switching. The aim of the remainder of this paper is to develop models that provide physical insight on the DW dynamics. The major simplification is to describe the wall by collective coordinates.

\section{Collective coordinate model} \label{CollectiveCoordModel}
\subsection{Choosing the collective coordinates} 
From the micromagnetic simulations we can infer that the dynamics comprises mainly the translation motion of a wall of fixed width but whose internal magnetization has a variable tilt. It is thus natural to choose the wall position $q$ as the main reaction coordinate along the reversal path. We assume for the sake of simplicity that the wall stays straight and of constant width. We thus first adapt the popular $\{q, \phi\}$ model~\cite{thiaville_domain_2004} to the disc geometry and use the tilt $\phi$ of the in-plane magnetization as the conjugate coordinate of the wall position $q$. In all calculations we will consider that the tilt is uniform (i.e. $\frac{\partial \phi}{\partial y}=0$ along the wall length). The validity of this assumption will be discussed in section \ref{beyondqphi}.
We fix a wall profile inducing a magnetization orientation with a polar angle being: 
\begin{equation}\theta (x,t)=2 \tan{^{-1}[ \exp{(\frac{x-q(t)}{\Delta}})]}.\label{dwprofile}\end{equation}
The wall dynamics can then be conveniently described in Lagrangian formalism.
\subsection{Equations of the domain wall motion within a disc} 
The Lagrangian $L=L_B-U_{\mathrm{tot}}$ of the system is the difference between the kinetic term: \begin{equation} L_B=\frac{M_s}{\gamma} \int_\textrm{disc} \Dot{\phi} (1-\cos{\theta})\,d^3r \end{equation}
and the total energy: 
\begin{equation}U_{\mathrm{tot}}=\int_\textrm{disc} u_{\mu mag}\, d^3r \end{equation}  
The total energy density $u_{\mu mag}$ includes the dipole-dipole interactions, the anisotropy energy density $u_{anis}= \frac{1}{2} \mu_0 H_ k M_s \sin^2{\theta}$, the exchange energy density $u_{ex}=A_\textrm{ex} (\frac{d\theta}{dx})^2$ and the Zeeman energy density $u_Z=- \mu_0 H_z M_s \cos{\theta}$ of the external field $H_z$. The dissipation function $W_G$ includes the effect of the damping $\alpha$ and the STT: 
\begin{equation}
    W_G= \frac{1}{2}\alpha \frac{M_s}{\gamma} \int_\textrm{disc} (\ \Dot{\theta}^2 + \Dot{\phi}^2 \sin^2{\theta})\,d^3r -  \sigma j \frac{M_s}{\gamma}\int_\textrm{disc} \Dot{\phi} \sin^2{\theta} \,d^3r
\end{equation}
 where $\sigma j$ is the STT expressed \cite{bouquin_size_2018} as:
 \begin{equation}
\sigma j=\gamma P \frac{V}{\mathcal{A}R_\perp}\frac{\hbar}{2e M_s t_\mathrm{mag}}
\end{equation}
The correspondence is $\sigma j = 1.1 \ \mathrm{GHz}$ for 0.85 V.

The wall dynamics is obtained from the Euler-Lagrange equations for $X=q,\phi$: 
\begin{equation}
    \frac{\mathrm{d}}{\mathrm{d}t}\frac{\partial L}{\partial \Dot{X}}-\frac{\partial L}{\partial X}+\frac{\partial W_G}{\partial \Dot{X}}=0
    \label{lagangian}
\end{equation}
Using the wall profile (Eq.~\ref{dwprofile}) and writing the volume integrals over the disc as $\int_\textrm{disc} d^3r = 2 d \int ^R _{-R} \mathrm{d}x \sqrt{R^2-x^2}$, the Euler-Lagrange equations can be simplified to the two following coupled differential equations: 
\begin{equation}
-\Dot{\phi} + \alpha \frac{\Dot{q}}{\Delta} = - \frac{\gamma}{2M_s d}\frac{\Delta}{S_\mathrm{DW}(q)}\frac{\partial U_\mathrm{tot}}{\partial q}
\label{model2a}
\end{equation}
\begin{equation}
\frac{\Dot{q}}{\Delta}+ \alpha \Dot{\phi}  = - \frac{\gamma}{2M_s d}\frac{1}{S_\mathrm{DW}(q)}\frac{\partial U_\mathrm{tot}}{\partial \phi} + \sigma j 
\label{model2b}
\end{equation}
where $S_\mathrm{DW}(q)$ has the dimension of a surface and reads:
\begin{equation}
S_\mathrm{DW}(q) = \int ^R _{-R} \mathrm{d}x \ \mathrm{sech}^2(\frac{x-q}{\Delta}) \sqrt{R^2-x^2}.
\label{sdw}
\end{equation}
$S_\mathrm{DW}(q)$ can be viewed as the effective surface of the DW, as it corresponds to the integral over the disc of $\sin^2(\theta)$, the in-plane projection of the magnetization. It can be reasonably approximated [Fig. \ref{Sdw}(b)] by the product of the wall width $\Delta$ by the chord length, yielding: 
\begin{equation}
S_\mathrm{DW}(q) \approx 2 \Delta \sqrt{R^2 - (R-q)^2}
\label{SDWapproximate}
\end{equation}

\subsection{Exact energy landscape for a disc geometry}
\begin{figure*}[t!]
\begin{center}
\includegraphics[width=15. cm]{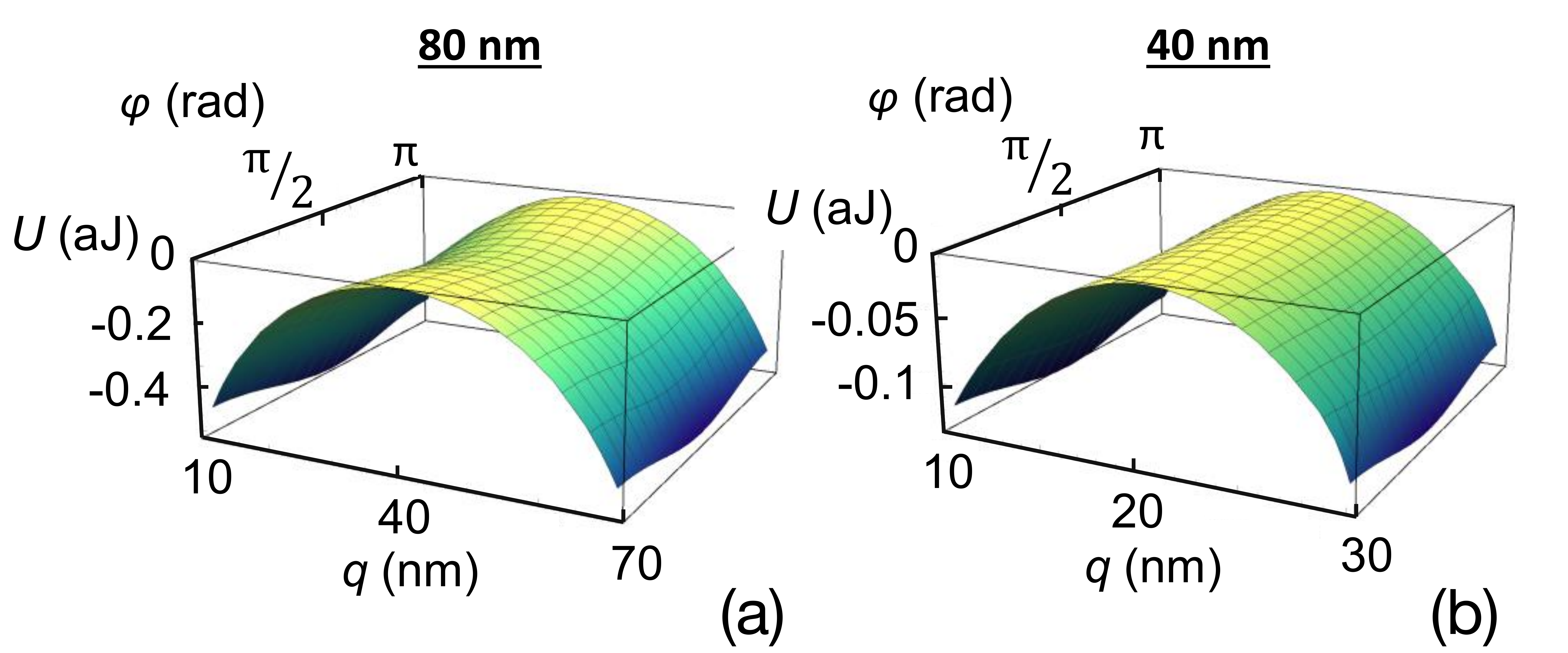}
\caption{Energy landscapes $U_\textrm{tot}$ extracted from micromagnetic simulations for all positions $q$ and tilt angles $\phi$ for a straight domain wall in perpendicularly-magnetized discs of diameters (a): 80 nm and (b): 40 nm. A constant offset was added to the energy landscape to have the energy maximum be zero.}
\label{figureUtot}
\end{center}
\end{figure*}

The main complication in Eq.~\ref{model2a}-\ref{model2b} resides in the evaluation of the energy $ U_\mathrm{tot}$ because it includes the dipole-dipole interactions that are non-local. To circumvent this difficulty, a first option is to use micromagnetic simulations to get $U_\mathrm{tot}(q,\phi)$ for all tilts and all DW positions inside the disc. Since the half domain wall width $\pi \Delta /2$ is 13 nm we restrict to wall positions at least 10 nm away from the edge. 
The energy landscapes (Fig.~\ref{figureUtot}) have saddle shapes with curvatures for the $q-\phi$ degrees of freedom of the DW.  
The energy is maximum when the wall is at the center of the disc and when the wall adopts in addition a N\'eel configuration ($\phi= 0~[\pi]$). This energy difference between N\'eel and Bloch configurations is much stronger for the largest diameter.

Before detailing the complicated situation of the disc, it is useful to recall the well-known DW dynamics in stripe \cite{cucchiara_domain_2012, thiaville_micromagnetic_2005, devolder_time-resolved_2016}. Indeed in a stripe the wall position is not tightened to the wall length (which is constant). The dynamics is much simpler but will shed light onto the forthcoming more complicated case of the disc.

\subsection{Domain wall motion in a stripe} 
\begin{figure}[!t] 
\begin{center}
\includegraphics[width=8 cm]{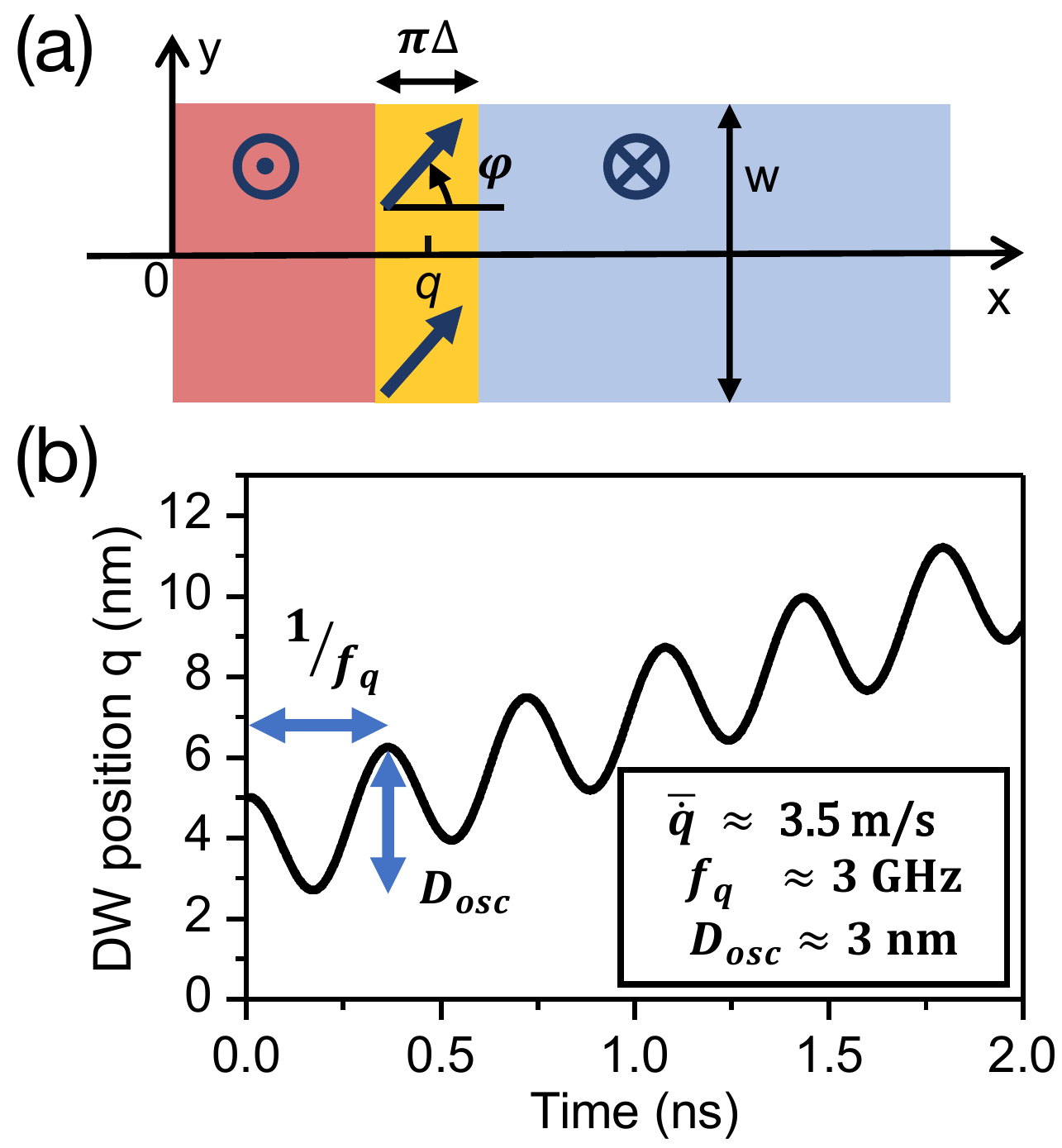}
\caption{Dynamics of a domain wall in an infinitely long stripe in the $q-\phi$ model. (a) Geometry: a straight wall (yellow) of width $\pi \Delta$ and of tilt $\phi$ is at a position $q$ within a stripe of width $w$. (b) Domain wall position as a function of time for a spin-torque $\sigma j = 0.5 \ \mathrm{GHz}$, an applied field $\mu_0 H_z= 50 \ \mathrm{mT}$ and material parameters modeling a CoFeB free layer with perpendicular anisotropy.}
\label{figurestripe}
\end{center}
\end{figure}    
 In an infinitely thin stripe of width $w$, the dipolar energy can be described as a \textit{local} quantity comprising two contributions. The demagnetizing field oriented in the out-of-plane direction can be grouped with the magneto-crystalline anisotropy into an effective anisotropy energy density $K_\textrm{eff} \sin^2{\theta}$, or an effective anisotropy field $H_k^\textrm{eff}=H_k-M_s$. The contribution from the in-plane components of the demagnetizing fields that appear within the wall can be described as a DW anisotropy energy density $u_{DW}$ \cite{mougin_domain_2007} which reflects the fact that N\'eel-type DWs have volume magnetic charges (i.e. we have $\vec{\nabla}. \vec{M} \neq 0$ within the entire volume of the wall) while Bloch-type DW have surface charges only at the two edges of the stripe. The DW anisotropy energy density reads $u_{DW}=K_{DW} \sin^2{\phi} \cos^2{\theta}$ where $K_{DW}=\mu_0 \frac{H_{N\leftrightarrow B} M_s}{2}$ with the DW stiffness field $H_{N\leftrightarrow B}$ being the in-plane field that one needs to apply to transform a Bloch DW into a N\'eel DW. (Note that $H_{N\leftrightarrow B}$ was written $H_{DW}$ in some previous papers~\cite{devolder_time-resolved_2016}).
 
 The DW stiffness field can be written \cite{mougin_domain_2007} from effective demagnetizing factors of the DW: $H_{N\leftrightarrow B}=\frac{M_s}{2}(N_y-N_x)$. If $w \gg d $ and $ w \gg \pi \Delta$ the demagnetizing factors are: 
 \begin{equation}
 N_x \approx \frac{d}{d+w} \textrm{~~and}~~  N_y \approx \frac{d}{d+\pi \Delta}    
 \label{demagfactors}
 \end{equation} 
 For example, a width $w=80 \ \mathrm{nm}$ and a thickness $d = 2 \ \mathrm{nm}$ yields a stiffness field of $\mu_0 H_{N\leftrightarrow B}=34 ~\textrm{mT}$.

With these local expressions of the energy terms, and using the volume integrals of the stripe $\int_\textrm{stripe} d^3r$ instead of that of the disc in the different terms of the Lagrangian, the equations of motion (Eq.~\ref{model2a}-\ref{model2b}) in a stripe become very simple:
\begin{eqnarray}
    &-\Dot{\phi} + \alpha \frac{\Dot{q}}{\Delta} = - \gamma_0 H_z\\
    &\frac{\Dot{q}}{\Delta}+ \alpha \Dot{\phi}  = \gamma_0 \frac{H_{N\leftrightarrow B}}{2} \sin{2\phi} + \sigma j
\label{modelstripe}
\end{eqnarray}
The nature of the DW motion depends on \cite{thiaville_micromagnetic_2005} the relative strengths of the applied field $H_z$ and the Walker field $H_\mathrm{Walker}=\alpha \frac{H_{N\leftrightarrow B}}{2}$. If $|H_z| \leq H_\mathrm{Walker}$ (i.e. at very small fields, typically less than 0.3 mT) the DW moves at constant velocity and constant tilt. In the alternative case the wall is in the precessional regime: $q$ and $\phi$ undergo coupled oscillations but in average the DW advances in the direction favored by the applied stimuli [Fig.~\ref{figurestripe}(b)]. 
It can be shown that: \begin{itemize}
    \item the mean DW drift velocity is $\bar{\Dot{q}} = \Delta (\sigma j - \alpha \gamma _0 H_z)$,
    \item  the tilt rotates at a rate $\omega_\phi = \gamma _0 H_z + \alpha \sigma j$,
    \item  the frequency of the position oscillations is $\omega_q = 2 |\omega_\phi|$,
    \item  the amplitude of the back-and-forth position oscillation is $D_\mathrm{osc} \approx \Delta \frac{H_{N\leftrightarrow B}}{2 H_z}$.   
\end{itemize}
We will see that the Walker field is much smaller that the effective fields to be at play in a disc (Fig.~\ref{Sdw}): the precessional regime of wall propagation will be the only one occurring in disc and the above dynamical features will be preserved to some extent. 

The result of this model calculation --rigid wall in an infinitely long stripe at zero temperature-- bears some similarity with the experimental behavior in the elongated rectangles at large stimulus [Fig.~\ref{wavy80nm}(a)] and at room temperature. The above itemized points shed light onto some of the experimental findings. For instance, small applied fields increase the frequency of the oscillation and reduce its amplitude, which explains why in experiments the oscillations can easily be hidden by the noise as soon as some field is applied. 

However this model calculation is not able to describe the behavior at small stimulus, with the propensity of making sometimes many oscillations at distributed intermediate conductance levels [Fig.~\ref{wavy80nm}(b)]. Understanding this feature requires to grab the behavior of a more realistic wall by taking into account the possibility of non-uniform tilt and presence of a partial Bloch line. This will be done in section~\ref{beyondqphi}.

\section{Approximate analytical field model of the STT-induced DW motion within a disc} 
\label{ApproximateAnalyticalSection}
We now return to case of the disc, with the objective of obtaining didactic expressions describing the dynamics. 
\subsection{Approximate analytical formulation of the forces acting on a straight wall in the collective coordinate model} 
It is possible (but cumbersome) to integrate directly the energies to get the generalized forces $\frac{\partial U_\mathrm{tot}}{\partial q}$ and $\frac{\partial U_\mathrm{tot}}{\partial \phi}$ and simplify the equations of motion (Eq.~\ref{model2a}-\ref{model2b}) to more didactic ones. However in order to shed light onto the different physical effects at play, we prefer to derive the equations of motion from an equivalent but more intuitive way by formally splitting the total energy in its Zeeman and stray field part $U_Z$, its part originating from the domain wall length $U_\mathrm{elastic}$ and its part $U_{\phi, NB}$ originating from the in-plane demagnetizing fields of the DW.

The Zeeman energy $U_Z$ is simply evaluated by the surface in which the magnetization has changed by $-2M_s$. Noticing that $\ell_\textrm{wall} \equiv S_\mathrm{DW}(q) / \Delta$ is the wall effective length when at position $q,$ the Zeeman energy can be written as:
$$
U_Z(q)=-\mu_0 d M_s H_z \left( \pi R^2 - 2 \int_0 ^q \frac{S_{DW}(x)}{\Delta} dx  \right)
$$ 
such that the Zeeman pressure acting on the wall is:
\begin{equation}
\frac{\partial U_Z}{\partial q} =  2 \mu_0 d M_s  \frac{S_{DW}(q)}{\Delta} H_z    \label{Uz}
\end{equation}
In analogy, the two domains at $x \geq q$ and $x \leq q$ create a dipole field $H_d(q)$ which is along $z$ at the wall position. The calculation of the dipole field is done in the supplementary material. $H_d(q)$ thus simply adds to $H_z$ in Eq.~\ref{Uz}.

As long as the wall is constrained to keep its native profile (Eq.~\ref{dwprofile}) the effective anisotropy energy and the exchange energy can simply be accounted for by multiplying the DW surface energy density $4 \sqrt{A_\mathrm{ex}K_\mathrm{eff}}$ by the wall cross-sectional area $d \ell_\textrm{wall}$ to get $U_\mathrm{el} = 4 d \sqrt{A_\mathrm{ex}K_\mathrm{eff}}  (S_\mathrm{DW}(q)/{\Delta})$. The elastic force acting on the wall is thus:
\begin{equation}
\frac{\partial U_{el}}{\partial q} = 4 d \sqrt{A_\mathrm{ex}K_\mathrm{eff}} \frac{\partial S_\mathrm{DW}(q)}{\partial q}
\label{Uelastic}
\end{equation}

\begin{figure}[htp!] 
\begin{center}
\includegraphics[width=8.2 cm]{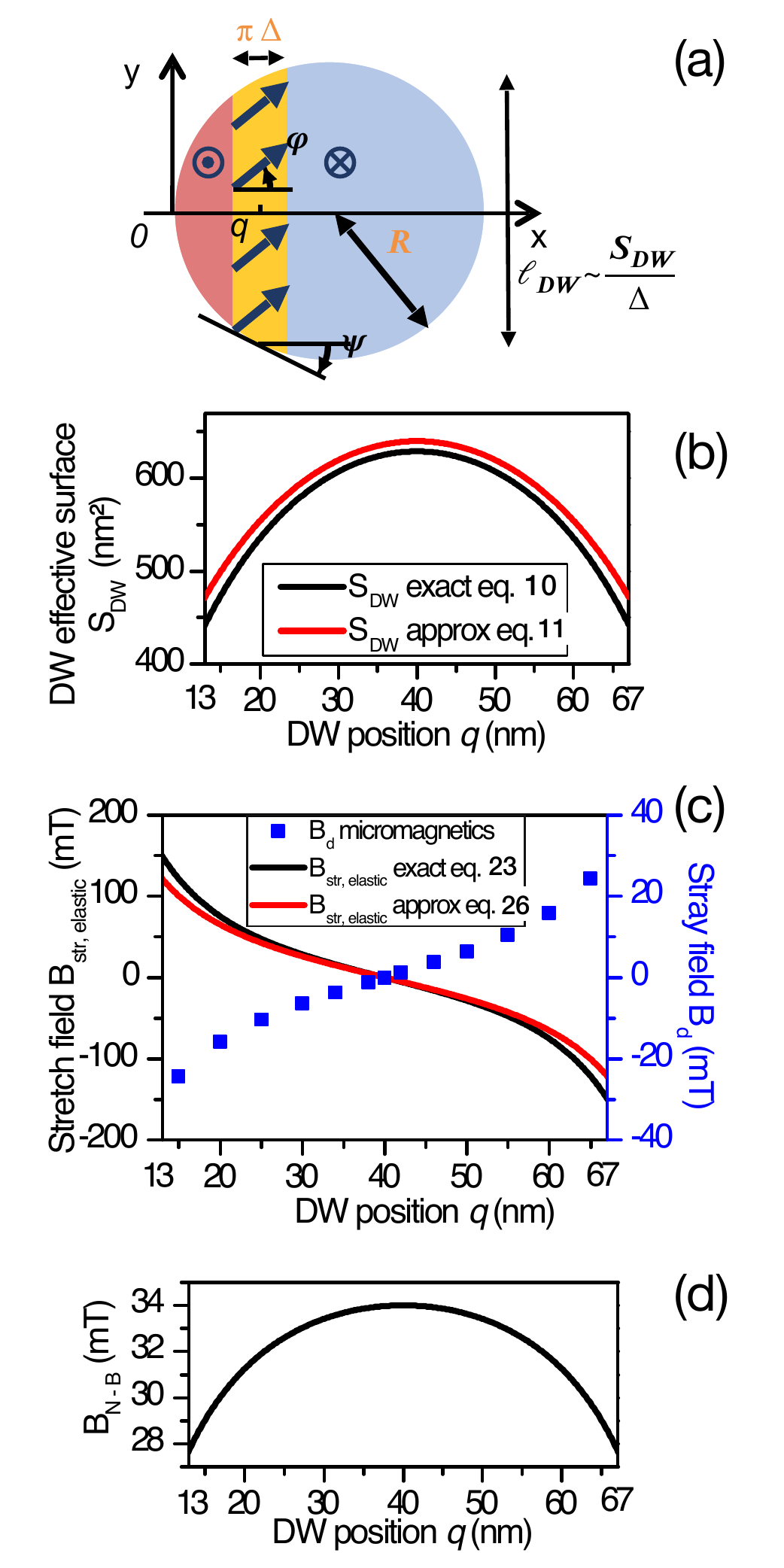}
\caption{(a) Sketch of the geometry and main definitions. (b) Effective surface $S_{DW}(q)$ of a straight domain wall in an 80 nm disc as computed exactly using Eq.~\ref{sdw} and as approximated using the product of the length of the chord by $\Delta$. (c) Elastic part of the stretch field (Eq.~\ref{stretchfield} and its large diameter approximation when confusing the wall length and the disc chord, as performed in Eq.~\ref{Hstrchord}) and stray field $H_d(q)$ of the two domains versus DW position in an 80 nm disc. (d) Domain wall stiffness field $H_{N\leftrightarrow B}(q)$ versus DW position in an 80 nm disc.
}
\label{Sdw}
\end{center}
\end{figure}

Finally, the energy arising from the in-plane demagnetizing fields of the wall can simply be written as the product of the effective volume in which there are volume charges $2 d S_\mathrm{DW}$ and its uniaxial N\'eel-Bloch anisotropy field $H_{N\leftrightarrow B}$, i.e. 
\begin{equation} U_\mathrm{\phi, NB} = \frac{1}{2}\mu_0 H_{N\leftrightarrow B} M_s \cos^2(\phi) \times (2d~ S_\mathrm{DW}(q))\label{UNBphi}\end{equation}
For simplicity we assume that $H_{N\leftrightarrow B}$ is independent from $q$. This approximation is of minor importance [see Fig.~\ref{Sdw}(d)]. The somewhat non-intuitive factor of 2 in the effective volume $2 d S_\mathrm{DW}$ of the magnetic charge distribution of a N\'eel wall was obtained in the exact derivation of $U_\mathrm{\phi, NB}$ and arises from the integration of the hyperbolic secante (not shown). 
This energy term yields two forces. The first force favors the N\'eel to Bloch transition and reads:
\begin{equation}
\frac{\partial U_{\phi, NB}}{\partial \phi} =- \mu_0  H_{N\leftrightarrow B} M_s d~ S_\mathrm{DW}(q) \sin(2 \phi)
\end{equation}
The second force favors the domain wall motion only when the wall has a N\'eel component (i.e. when $\phi \neq \pi/2~[\pi]$) and reads:
\begin{equation}
\frac{\partial U_{\phi, NB}}{\partial q} = \mu_0  H_{N\leftrightarrow B} M_s d \frac{\partial S_\mathrm{DW}(q)}{\partial q} \cos^2 (\phi)
\end{equation}

Summing all energies, the equations of motion become:
\begin{equation}
-\Dot{\phi} + \alpha \frac{\Dot{q}}{\Delta} = - \gamma_0 \left[H_z+H_\mathrm{d}(q)+H_\mathrm{str}(q, \phi)\right]
\label{model3a}
\end{equation}
\begin{equation}
\frac{\Dot{q}}{\Delta}+ \alpha \Dot{\phi}  = \gamma_0 \frac{H_{N\leftrightarrow B}}{2} \sin{2\phi} + \sigma j, 
\label{model3b}
\end{equation}
where we have defined the \textit{domain wall stretch} field $H_\mathrm{str}(q, \phi)$ whose physical meaning is discussed in the following section. 

\subsection{Main driving force: the stretch field} 
The stretch field $H_\mathrm{str}(q, \phi)$ contains two contributions: an \textit{elasticity} part $H_\mathrm{str, el}(q)$ that solely depends on the wall position, and a \textit{dipolar} part $H_\mathrm{str,~NB}(q, \phi)$ that depends on both $q$ and $\phi$ and that is related to the N\'eel or Bloch (NB) nature of the domain wall.\\

The elasticity part of the stretch field reads:
\begin{equation} H_\mathrm{str,~el}(q) \equiv \frac{2 \sqrt{A_\mathrm{ex}K_\mathrm{eff}}}{\mu _0 M_s} \frac{1}{S_\mathrm{DW}(q)} \frac{\partial S_\mathrm{DW}(q)}{\partial q} \end{equation}
or equivalently: 
\begin{equation}
H_\mathrm{str,~el}(q) = H_k^\textrm{eff}\frac{\Delta}{S_\mathrm{DW}(q)} \frac{\partial S_\mathrm{DW}(q)}{\partial q}
\label{stretchfield}
\end{equation}
The elasticity part of the stretch field simply illustrates that while moving within the disc, the length of the wall varies and this costs or provides energy. $H_\mathrm{str,el}$ plays the same role as an out-of-plane external field: it applies a pressure on the wall. 

The other part of the stretch field depends both on the wall position $q$ and on the wall configuration $\phi$ (N\'eel or Bloch, NB). This $ H_\mathrm{str,NB}$ accounts for the fact that when a N\'eel DW is displaced, this induces a change of the volume in which this N\'eel wall induces volume magnetic charges. It reads:
\begin{equation}
     H_\mathrm{str,NB}(q, \phi)\equiv \frac{H_{N\leftrightarrow B}}{2} \frac{\Delta}{S_\mathrm{DW}(q)} \frac{\partial S_\mathrm{DW}(q)}{\partial q}\cos^2{\phi}
     \label{NBstretchfield}
\end{equation}
From Eq.~\ref{stretchfield} and \ref{NBstretchfield}, it is clear that the spatial variations of the two parts of the stretch field are identical. However they have very different magnitudes, with ratio:
\begin{equation} \frac{H_\mathrm{str, NB}}{H_\mathrm{str, el}} \leq \frac{H_{N\leftrightarrow B}}{2H_k^\textrm{eff}} \label{coco} \end{equation}
so that the dipole-dipole part of the stretch field is typically a tenth of the elasticity part.

For large discs with $R \gg \Delta $, the stretch field can be written in a simpler manner as: 
\begin{equation}
    H_\mathrm{str}(q, \phi, R \gg \Delta) \approx \left( H_k^\textrm{eff} + \frac{1}{2}H_{N\leftrightarrow B}  \cos^2\phi\right)\frac{\Delta (R-q)}{q(2R-q)}
\label{Hstrchord}
\end{equation}
Note that near the disc center, the part of the curvature of the total energy that is related to the stretch field follows $ \frac{\partial H_\mathrm{str}}{\partial q} \propto - \frac{\Delta}{R^2}$. This quadratic decrease with the inverse diameter indicates that the stretch field at the center of the disc is a relevant parameter only for relatively small discs.

As a side remark, we mention that the concept of stretch field is not restricted to our specific geometry. If a straight wall was hypothetically moving in a funnel of angular opening $2\psi$ [see Fig.\ref{Sdw}(a)], the stretch field could be rewritten as:
\begin{equation}
H_\mathrm{str}(q,\phi) = \left({2 H_\textrm{k,eff} + H_{N\leftrightarrow B}\cos^2(\phi)} \right) \frac{\Delta}{\ell_{wall}(q)} \tan(\psi(q))
\label{stretchfieldpsi}
\end{equation}
where $\ell_{wall} \equiv S_\mathrm{DW}(q) / \Delta$ would still be the wall length. The $\tan[\psi(q)]$ term in Eq.~\ref{stretchfieldpsi} makes the stretch field very sensitive to the exact shape of the device at the location of the wall. This is particularly critical at the center of the device. For instance if the nominally rectangular devices (i.e. $\sqsubset \! \sqsupset$) are distorted because of some miscorrection of the lithography proximity effect, the device shape might either be stadium-like (i.e. $\subset \! \supset$), leading to a maximum of energy at the center, either bowtie-like (i.e. $\rhd  \!\!  \lhd$) leading to a secondary minimum of energy. The sign of the stretch field is reversed between these two situations. In the bowtie-like distorsion, the wall oscillations can be bound to the center of the device, as commonly practiced to get artificial wall pinning at notches along stripes.

\subsection{Amplitudes of the different disc-specific effective fields} 

The numerical evaluations of the effective fields are reported in Fig.~\ref{Sdw}c and d, and further detailed in the supplementary material. The stretch field is by far the largest relevant field during the DW motion. For instance for the disc with $2R = 80 \ \mathrm{nm}$  [Fig.~\ref{Sdw}(c)] the elastic part of the stretch field reaches $\pm150 \ \mathrm{mT}$ when the DW is near the edges. Its (smaller, see Eq.~\ref{coco}) demagnetizing part does not exceed 35 mT [Fig. \ref{Sdw}(d)].

The stray field $H_\mathrm{d}(q)$ acting on the DW depends on the wall position and needs to be evaluated by micromagnetics. 
This field is zero when the two domains are of equal size. It is maximum when the wall is at an edge. The stray field in a 80 nm disc [Fig. \ref{Sdw}(c)] is typically 5 times smaller and of opposite sign than the stretch field. It has a shape that resembles very much that of the elasticity part of the stretch field. This shape similarity is discussed in the supplementary material.

\section{Comparison of the predictions of the different models}  \label{SectionComparaisonDesModeles}
We have described so far the exact dynamics of a wall as well as a simplified formalism by assuming that the micromagnetic state can be described by the sole wall position and tilt. This collective coordinate model can either be solved exactly based on the numerical evaluation of the total energy (model to be quoted as the $q-\phi-U_{tot}$ model, Eq.~\ref{model2a}-\ref{model2b}) or it can be solved using the approximate analytical expressions of the effective fields (hereafter quoted as the $q-\phi-H_{str}$ model, Eq.~\ref{model3a}-\ref{model3b}, section~\ref{ApproximateAnalyticalSection}). The goal of this section is to assess the level of accuracy of the collective coordinate models. 
\subsection{Comparison of the qualitative features predicted by the different models}

\begin{figure}[t!]
\begin{center}
\includegraphics[width=8.5 cm]{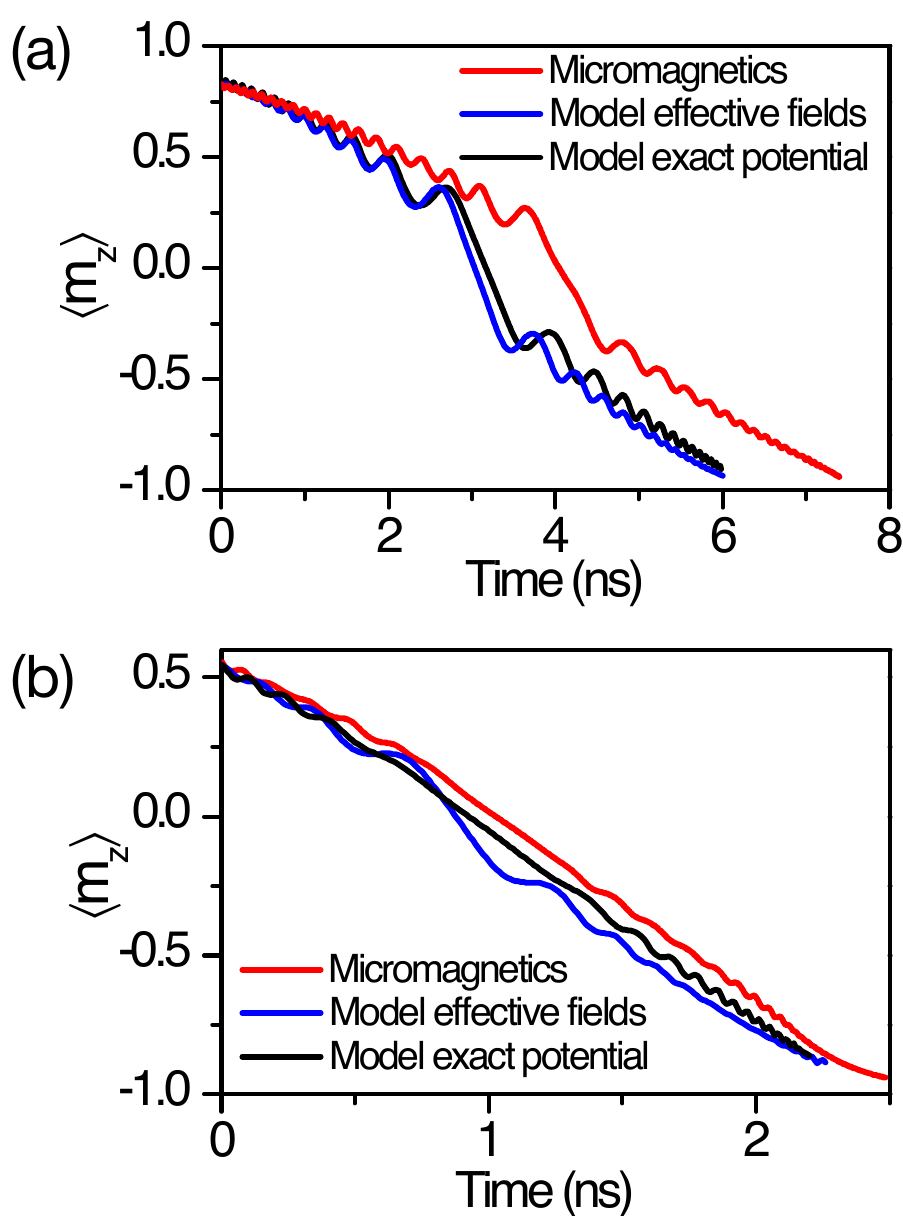}
\caption{Comparison between the micromagnetic simulation of the dynamics of a wall with initial optimal curvature (red curves), the collective coordinate model with the exact energy landscape ($\{q,\phi\}$ model with exact $U_\textrm{tot}$, black curves), and with the approximate effective fields (blue curves). a) Mean value of the $z$ component of the magnetization evolution during the DW motion for a disc of 80 nm diameter. The initial conditions are for $q_0=10~\textrm{nm}$ and $\phi_0=70~\textrm{deg.}$.  b) Idem for a diameter of 40 nm. \label{quantitativecomparison}
} \end{center}  \end{figure}

Figure ~\ref{quantitativecomparison} compares the time-resolved magnetic moments deduced from the three models. Figure~\ref{comparisonfrequencies} focuses on the metrics of the oscillatory part of the dynamics. Figure~\ref{qualitativecomparison} illustrates the sensitivity to the initial conditions within the $q-\phi-H_{str}$ model.

The main features formerly identified in the micromagnetic model -- the DW drift motion, its superimposed oscillation that get more pronounced and slower when the DW is near the disc center-- are qualitatively reproduced in the $q-\phi$ models. The exact and the approximate $q-\phi$ models yields very similar domain wall dynamics within a 80-nm disc; the sensitivity to the initial conditions is also reproduced. For a 40 nm diameter, there is a good agreement between the micromagnetic model and the $q-\phi-U_{tot}$ model but the approximate $q-\phi-H_{stretch}$ model is not as satisfactory, which reveals the limits of our analytical approximations at small disc diameters.

These main qualitative features can be understood from the equations of motion of the $q-\phi-H_{str}$ model (Eq.~\ref{model3a}-\ref{model3b}) because they resemble that of the stripe, except that the field part of the DW driving force $H_\mathrm{tot}(q)=H_z + H_\mathrm{str} + H_\mathrm{d}$ is now position dependent. Except in a very small position interval which is at the center of the disc when $H_z=0$, this total field exceeds the Walker field such that the wall motion is still occurring in the precessional regime. In analogy with the stripe, the drift part of the DW motion can be written from an average velocity: 
\begin{equation} \bar{\Dot{q}} \approx \Delta \left[\sigma j - \alpha \gamma _0 H_\mathrm{tot}(q)\right],~\mathrm{provided}~ H_\mathrm{tot}(q) \neq 0
\label{driftvelocity}
\end{equation}
We expect oscillations of the DW position being: \begin{equation} D_\mathrm{osc} \approx \Delta \left| \frac{H_{N\leftrightarrow B}}{2 H_\mathrm{tot}(q)} \right| ,~\mathrm{provided}~ H_\mathrm{tot}(q) \neq 0.
\label{Dosc}
\end{equation} Still in the same analogy, the tilt is also expected to rotate at the rate:
\begin{equation} \omega_\phi  \approx \gamma _0 H_\mathrm{tot}(q) + \alpha \sigma j,~~\mathrm{provided}~ H_\mathrm{tot}(q) \neq 0  \label{omegaq}\end{equation}

\subsection{Quantitative comparison of the domain wall drift velocity and oscillation periods within of the different models}

\begin{figure}[t!] 
\begin{center}
\includegraphics[width=8.5 cm]{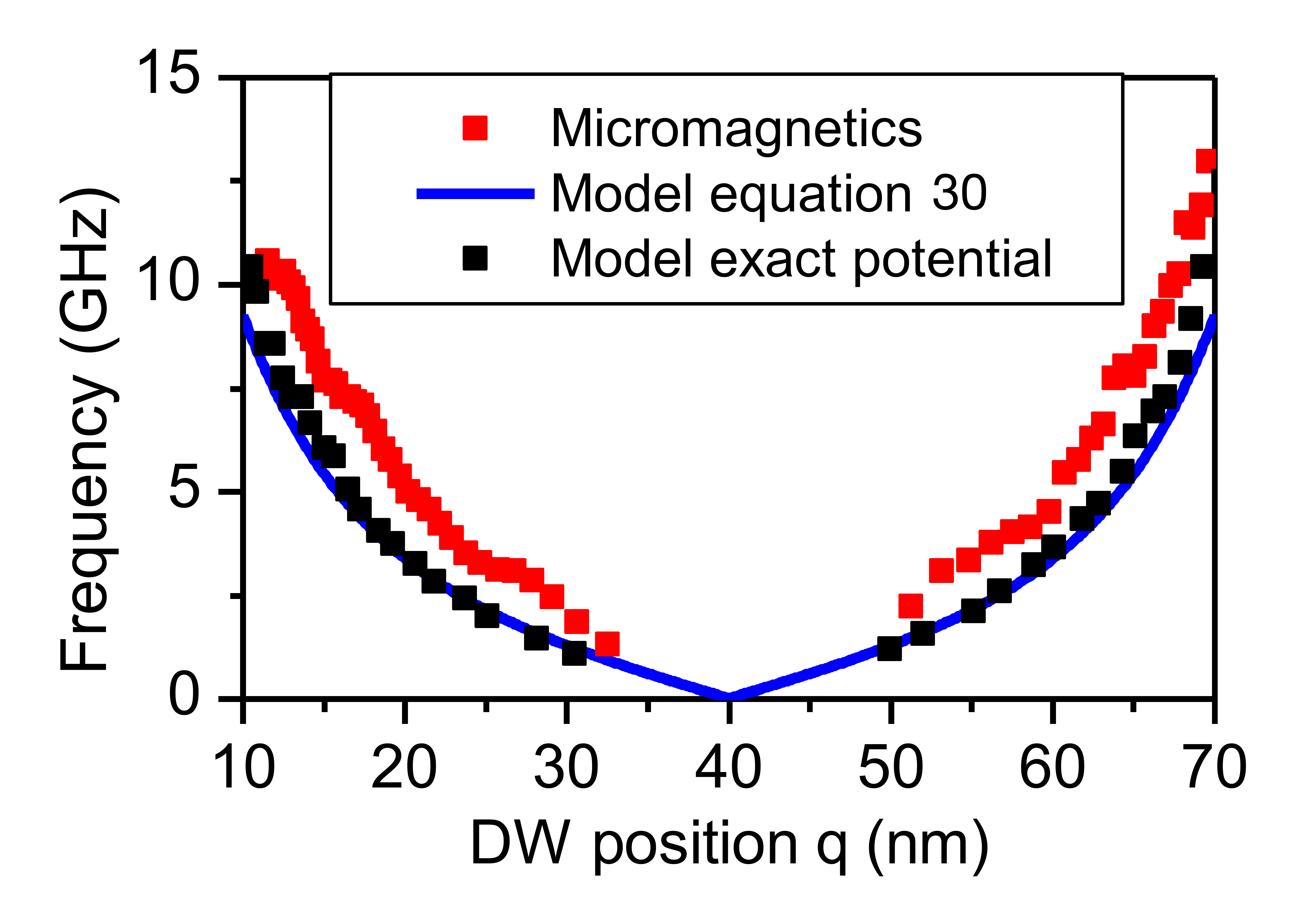}
\caption{Absolute value of the frequency of the oscillatory part of the domain wall motion in a disc of diameter 80 nm. Relation between the inverse of the wall oscillation period and the mean wall position during this period as predicted by  micromagnetics (red symbols) and by the $q-\phi-U_{tot}$ model (black symbols). The blue line is frequency inferred by analogy with the infinite stripe (Eq.~\ref{omegaq}). The frequencies change sign at the middle of the disc.} \label{comparisonfrequencies} \end{center} \end{figure}

\begin{figure*}[t!] 
\begin{center}
\includegraphics[width=16 cm]{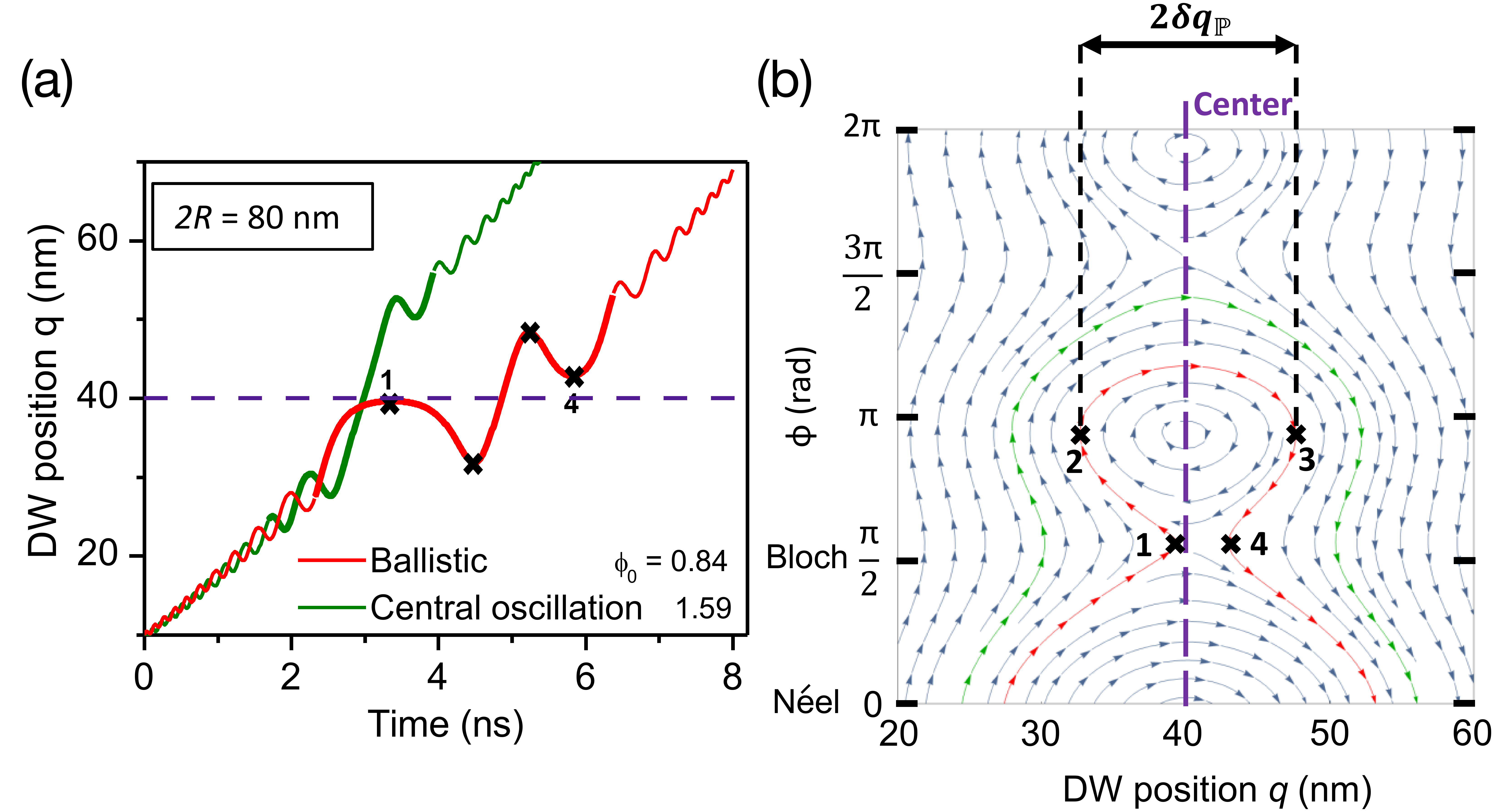}
\caption{Domain wall dynamics within a disc of 80 nm diameter under STT in the collective coordinate model with approximate effective fields. a) Time dependence of the position of the DW, for two different initial tilt angles being 48 deg. and 91 deg. and the same initial positions $q_0=10~\textrm{nm}$. 
b) Description of the possible trajectories of the DW in the space of the collective coordinates. $\phi=0$ corresponds to a N\'eel DW and $\phi = \pi /2$ to a Bloch DW and the landscape is $\pi$ periodic in the $\phi$ direction. The colored trajectories correspond to the bold-colored parts of the trajectories of panel (a). The contour between the labels 1,2,3 and 4 of the red trajectory is very close to the retention pond frontier $\partial \mathbb{P}$.}
\label{qualitativecomparison}
\end{center}
\end{figure*}

From the comparisons done in Figs.~\ref{quantitativecomparison} and ~\ref{comparisonfrequencies} we conclude that $q-\phi$ models predict correctly the DW drift velocities while they are much less accurate to account for the oscillatory part of the motion. This can be understood with the help of Eq.~\ref{driftvelocity} and~\ref{omegaq}. Indeed $\sigma j$ is a constant number ($\approx 1$ GHz) during the wall motion. The term $\gamma_0 H_\mathrm{tot}(q)$ is larger and position dependent. It varies between -25 and 25 GHz from one edge to the other in a 80-nm disc [Fig.~\ref{Sdw}(c)]. 

Having in mind the typical values of $\gamma_0 H_\mathrm{tot}(q)$ and  $\sigma j$, we find that the term $\alpha \sigma j$ is almost everywhere much smaller than $\gamma_0 H_\mathrm{tot}(q)$, such that the dynamics of the tilt (Eq. \ref{omegaq}) is resulting from the sole effective fields. Since the effective fields depend on the exact wall shape and exact tilt state, their accuracy is directly endangered by the approximation of a straight wall with uniform tilt; the frequency mismatch between the predictions of micromagnetics and of the $q-\phi$ models can thus be interpreted as a failure to account precisely enough for the stretch field.

The situation is reversed for the DW drift velocity: the term $\alpha \gamma_0 H_\mathrm{tot}(q)$ amounts at best to a minor fraction (e.g. circa 25\% when $2R=80$ nm) of $\sigma j$. The drift velocity prediction (Eq.~\ref{driftvelocity}) is thus little affected by potential errors in the stretch field. Besides, the term $\sigma j$ is independent of the wall fine structure, hence it is reliable despite our straight wall and uniform tilt approximations. This explains why the predictions of the DW drift dynamics are satisfactory.
Let us thus now use the collective coordinate models to discuss the sensitivity of the wall dynamics to the initial conditions.

\section{Origin of the sensitivity of the dynamics to initial conditions for large discs} \label{SectionSensitivityToInitialConditions} 
The sensitivity of the wall dynamics to the initial conditions for the largest disc, and the absence thereof for the smallest disc is best understood when analyzing  the DW trajectories in the $(q,\phi)$ space [Fig.~\ref{qualitativecomparison}(b)]. A convenient preliminary step is to discuss a \textit{gedanken experiment}: What would happen if a N\'eel wall was placed at the center of the disc ?
\subsection{Preamble: evolution of an hypothetical central N\'eel wall} 
Let's thus place a straight N\'eel wall at $\{q_0, \phi_0 \}=\{R, 0\}$ i.e. in the absolute maximum of the energy landscape. Some work of non-conservative torques is needed for the wall to move away from this energy maximum. In the $\{q, \phi \}$ space, the resulting trajectories are outgoing spirals: the wall first swings about the center with an increasingly large amplitude. In this transient process, there is a back-and-forth transfer of energy between the position degree of freedom of the wall and the N\'eel/Bloch degree of freedom of the wall: $U_{el}$ and $U_{NB,\phi}$ act as communicating vessels. 

This zone of the $\{q,\phi\}$ space in which the wall is transiently stuck is a "retention pond" \cite{bouquin_stochastic_2021}, written $\mathbb{P}$ and illustrated as the red contour from labels 1 to 4 in Fig.~\ref{qualitativecomparison}(b). As a side remark, let us note that this concept of "retention pond" is also implicitly present in the independent study presented in the Fig. 4 of ref. \cite{statuto_micromagnetic_2021}. Within the $q-\phi$ model, all trajectories crossing the frontier $\partial\mathbb{P}$ are \textit{outgoing} trajectories. In the absence of STT, $\mathbb P$ is centered on $\{q=R, \phi =0~[\pi]\}$. It is distorted for strong values of the STT (see supplementary material). As a result a DW that exits from $\mathbb P$ will subsequently cross the disc center but it never performs so while being in a N\'eel configuration, in line with the conclusions of the micromagnetics study for this diameter of 80 nm (section~\ref{sectionmumag}). Note that for discs smaller than a threshold to be determined later (Eq.~\ref{diameterForNoPond}), $H_{N\leftrightarrow B}$ is negative and $\mathbb P$ is then centered about the Bloch configuration instead of the N\'eel one. Extrapolation of our arguments to that case is trivial.

\subsection{Sensitivity of the dynamics to the initial conditions within $\{q-\phi\}$ model}
Let us return back to our case study which is the wall dynamics after an hypothetical nucleation of a DW near the disc edge, i.e. out of the retention pond.
The sensitivity of the dynamics to the initial conditions can be understood as follows. The DW progressive drift and the superimposed DW oscillations are rather independent phenomena such that when wall heads to the center of the disc, it can either approach close to the retention pond $\mathbb{P}$ or pass at a large distance from it. 
If the wall avoids the vicinity of $\mathbb{P}$, it crosses the disc center and performs a one-way, single attempt (ballistic) crossing. If in contrast the wall happens to approach the retention pond, it will have to circumvent it which leads to a more complex trajectory.

These two scenarios are illustrated in Fig.~\ref{qualitativecomparison}. Along green trajectory ($\phi_0=48~\textrm{deg.}$), the wall crosses the disc center with a quasi-Bloch configuration, far from the retention pond. This trajectory is archetypal of the ballistic crossing and most probably corresponds in experiments to the featureless and ramp-like conductance waveforms [Fig.\ref{wavy100nm}(e)].
If adding $\Delta \phi_0 \approx \pi/4$ to the initial tilt (red curve in Fig.~\ref{qualitativecomparison}), one obtains the antonym case: the DW trajectory arrives in a tangent manner to (but outside) $\mathbb{P}$. The wall performs a single turn about $\mathbb{P}$ thereby making a considerable back-and-forth motion with two pauses at either sides at the disc center.  This most probably corresponds to the experimental waveforms that exhibited a pronounced oscillation near the midway conductance [Fig.\ref{wavy100nm}(f)].

In short, the $\{q-\phi\}$ models can explain \textit{some} sensitivity to the initial conditions as depending $\{q_0,\phi_0\}$, the crossing will be either of the type drift-like one-way "ballistic" crossing, or will entail pre- and post-crossing pauses and a central oscillation. However the (more rarely occurring) swing-like crossings with multiple attempts [simulations of Fig.~\ref{LLGS80nm}(d), likely corresponding to the experiments of Fig.~\ref{wavy100nm}(g)] cannot be obtained in the framework of the $\{q-\phi\}$ models. We will discuss this case later in section~\ref{beyondqphi}.

\subsection{Size of the retention pond verus disc diameter} \label{sizepondsection}
The probabilities of occurrence of the ballistic crossing and the one-way crossing with pre- and post-crossing pauses are correlated with the size of $\mathbb{P}$. If one aims at a reproducible DW propagation duration while not being able to control to the initial wall position and tilt (or if the thermal noise is large enough to let $q$ and $\phi$ diffuse with time), it is important to determine the size of $\mathbb{P}$. We define $\delta q_\mathbb{P}$ the half width of the retention pond in the $q$ direction [see Fig.\ref{qualitativecomparison}(b)] and derive it in the conservative limit. $\delta q_\mathbb{P}$ can be evaluated by comparing the energy of the saddle point of $U_\textrm{tot}$ with that of a N\'eel wall $\delta q_\mathbb{P}$ away from the center [these two states are close to the labels 2 and 1 in Fig.~\ref{qualitativecomparison}(b)]:

\begin{equation}
\left( U_{\phi, NB} +U_{el} + U_d \right)  \Bigg\rvert_{
    {\begin{matrix}
    _{\hspace{-5mm}\phi=\pi}\\
    _{q=R-\delta q_\mathbb P}\\
    \end{matrix} }}
= \hspace{-0mm}
\left( U_{\phi, NB} +U_{el} + U_d \right)  \Bigg\rvert_{
    {\begin{matrix}
    _{\hspace{-0mm}\phi=\frac{\pi}{2}}\\
    _{q=R}\\
    \end{matrix} }}
\label{pondcontourequation}
\end{equation}
If we approximate $\ell_{wall}$ as the length of the disc chord and if we neglect the dipole energy $U_d$, The half-size of the retention pond is:
\begin{equation}
    \delta q_\mathbb{P} \approx R \sqrt{\frac{H_{N\leftrightarrow B}}{H_\textrm{k,eff}}}
    \label{pondwidth}
\end{equation}
This leads $\delta q_\mathbb{P} \approx 11~\textrm{nm}$ for a disc of diameter 80 nm, in agreement with the numerical result in the absence of spin-torque (Fig. S3, supplementary material) as well as at the levels of spin-torque used for switching [Fig.~\ref{qualitativecomparison}(b)].

\subsection{Disc diameter for optimal reproducibility of the duration of the domain wall propagation}
Let us determine the diameter $d_{\nexists \mathbb{P}}$ for which the retention pond disappears. This situation happens when a wall placed at the center of the disc has the same energy when either in the N\'eel or in the Bloch state, i.e. when: 
\begin{equation}
U_{\phi,NB}(q=R, \phi=0)=U_{\phi,NB}(q=R, \phi=\frac{\pi}{2})
\label{diameterForNoPond}
\end{equation}
or equivalently when $H_{N\leftrightarrow B}=0$. An estimation of $d_{\nexists \mathbb{P}}$ from the approximate demagnetizing factors of the wall (eqs.~\ref{demagfactors}) is unfortunately bound to fail as these expressions do not hold in the obtained diameter range. Micromagnetics must be used to find the disc diameter leading to the degeneracy of centered Bloch and N\'eel walls.
With our material parameters the pond reduces to a single point when $d_{\nexists \mathbb{P}}=40~\textrm{nm}$, which was to be anticipating from the vanishing curvature $\frac{\partial^2 U_{tot}}{\partial \phi^2}$ at $q=R$ at this specific diameter [see Fig.~\ref{figureUtot}(b)]. 
The disc diameter $d_{\nexists \mathbb{P}}$ is particularly interesting. Indeed it ensures that all the DW trajectories pass through the disc center in a ballistic way: the walls cross the center directly and almost independently from their Bloch or N\'eel character (see Fig. S4 in supplementary material for an exhaustive description of all the trajectories). This recalls the conclusions of micromagnetics [Fig.~\ref{LLGS40nm}(b)] when the DW dynamics in a 40 nm disc was almost insensitive to the initial tilt. This also correlates with the linear aspect of the time-resolved magnetization curves at the crossing of the disc center for this diameter. We can expect to maximize the reproducibility of the time needed for a wall to sweep across a disc, a feature that is of great interest for spin-transfer-torque magnetic random access memories (STT-MRAM).

As a side remark, we mention that for diameters $2R <d_{\nexists \mathbb{P}}$, we have $H_{N\leftrightarrow B}<0$. As a consequence, while the retention pond formerly centered about the N\'eel state has disappeared, another one emerges from the central Bloch situation. However this new pond has little practical relevance with our material parameters. Indeed the energy difference between the Bloch and N\'eel states of a central wall for $2R <d_{\nexists \mathbb{P}}$ stays very small, of the same order as the thermal energy $k_B T \approx 4~\textrm{zJ} $ (25 meV) at room temperature. For comparison, the energy difference between the Bloch and N\'eel states for a central wall at 2R=80 nm was 44 zJ (275 meV).

\begin{figure*}[t!] 
\begin{center}
\includegraphics[width=14 cm]{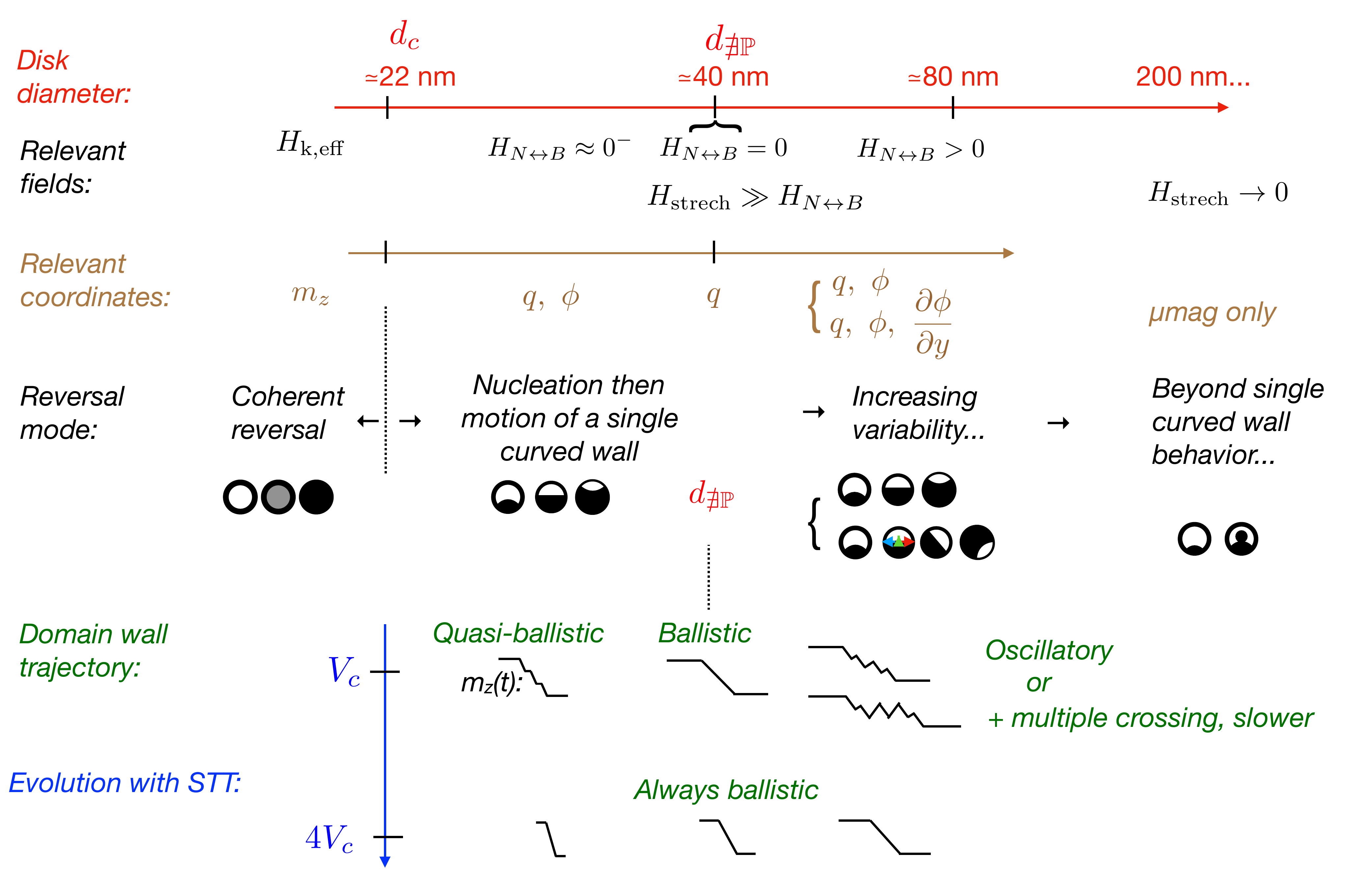}
\caption{Summarized description of the main features of the domain wall dynamics induced by spin-torque in perpendicularly magnetized discs. The disc diameter $d_{\nexists \mathbb P}$ is the one for which the position degree of freedom and the tilt degree of freedom of the domain wall get maximally independent when the wall crosses the disc center. The numerical values are for the material parameters of the theoretical sections. The conclusions of ref.~\cite{bouquin_size_2018} are also included.}
\label{FigSummary}
\end{center}
\end{figure*} 

\subsection{Domain wall dynamics beyond the $\{q-\phi\}$ models} 
\label{beyondqphi}
Let us come back to the 80 nm discs in which the micromagnetic simulations evidenced reversal paths that included multiple crossing of the disc center [Fig~\ref{LLGS80nm}(d)]. These switching paths cannot be described within the $q-\phi$ model; their understanding requires to take into account additional degrees of freedom. For these events, the micromagnetic configurations indicates that there is coincidence of two phenomena: the onset of a strong non-uniformity of the tilt ($\frac{\partial \phi}{\partial y} \neq 0$) and the trapping of the oscillating wall within the central part of the disc. Qualitatively, when the tilt is strongly non-uniform, different parts of the wall may want to move in opposite directions (see Eq.~\ref{modelstripe}), such that the wall ceases to advance but gyrates instead. 

The time-resolved system energy (not shown) indicates that during these swing-like scenarios the system systematically reaches a higher energy compared to when in the single crossing scenarios. For instance during the scenario of Fig~\ref{LLGS80nm}(d), the system energy rises 130 zJ (810 meV) above the maximum energy reached in Fig~\ref{LLGS80nm}(a). 
This extra energy cost can be understood from the theoretical energy \cite{hubert_magnetic_2008} of a full vertical Bloch line: $U_\textrm{B-line}= \frac{8A_\mathrm{ex} d}{\sqrt{H_k/M_s}} \approx 280~\textrm{zJ}$ (1.75 eV). Note that the expected size of a Bloch line ($\pi \sqrt{\frac{2A}{\mu_0 M_s^2}} \approx 15~\textrm{nm}$) makes it small enough to easily fit within a centered wall. 

To anticipate whether the tilt is likely to stay quasi-uniform during the DW motion, $U_\textrm{B-line}$ should be compared to the "DW energy barrier" $\Delta E_\textrm{dw} (q_0, \phi_0)$ that an (already nucleated) DW with uniform tilt needs to acquire to climb up to the saddle point in the energy landscape (i.e. the height of Fig.~\ref{figureUtot}). Since $\Delta E_\textrm{dw}$ depends on the subjectively chosen initial conditions $\{q_0, \phi_0\}$, only its order of magnitude is meaningful.
For 40 nm discs, the STT has to supply typically $\Delta E_\textrm{dw} \approx 120~\textrm{zJ}$ (0.75 eV) to jump over the barrier [Fig.~\ref{figureUtot}(b)]. This is much below $U_\textrm{B-line}$. With these energy scales in mind, we can understand why at diameters of 40 nm, the tilt is staying essentially uniform whatever the initial conditions $\{q_0,\phi_0\}$. Indeed it seems unlikely that minor changes in the reversal paths would permit a fourfold increase of the work supplied by the STT with the whole of it fed into the $\frac{\partial \phi}{\partial y}$ degree of freedom. This is why at diameters of 40 nm the $q-\phi$ model succeeds in describing the full wealth of the domain wall propagation: the energy cost of a strongly non-uniform $\phi$ is just out of reach.

Conversely for larger discs, the work supplied by the STT is substantially larger (e.g. $E_\textrm{dw} \approx 450~\textrm{zJ}$ (2.8 eV) for a diameter of 80 nm and $q_0=10~\textrm{nm}$). Minor changes in the reversal path and thus in the work supplied by the spin-torque may transfer a sizable part of this work in the $\frac{\partial \phi}{\partial y}$ degree of freedom of the wall. This can induce a swing-like scenario if the wall stays long enough at the disc center for a partial Bloch line to develop. 
In the case of the elongated rectangles, such swing-like scenario are considerably more probable. When in experiments, the wall seems to stop drifting during 3-4 oscillations [Fig.~1(b)], it is likely to result from the nucleation of vertical Bloch lines, that could enter the wall from the device edge at any random wall position.

\section{Discussion} 
\label{sectiondiscussion}
In this last section we discuss briefly some practical consequences that result directly from our modeling.
\subsection{Expected influence of an external field} 
Let us first examine the influence of an external out-of-plane field $H_z$ as this situation is frequently encountered in STT-MRAM applications in which offset fields emanating from other magnetic layers reach the free layer.  
These offset fields are generally below 20 mT, i.e. smaller \cite{devolder_offset_2019} than the typical value of the stretch field in discs; as a result they do not radically change the wall dynamics. An external field adds a gradient to the total energy $U_{tot}$ in the $q$ direction, and thus displaces the wall position $q$ for which the total effective field  $H_z+H_\mathrm{str}(q)+H_\mathrm{d}(q)$ vanishes. Using Eq.~\ref{Hstrchord} with $q\approx R$ and neglecting $H_d$ against $H_\textrm{stretch}$, we find that small applied fields lead to the displacement of the retention pond in the $q$ direction by approximately: 
\begin{equation}
\frac{H_z}{H_k^\textrm{eff}} \times \frac{R^2}{\Delta}
\label{pondshift}
\end{equation}
Some exact trajectories are reported in Fig.~S4 of the supplementary material. An offset field of 30 mT displaces the retention pond by 10 nm in disc of 80 nm of diameter, in agreement with Eq.~\ref{pondshift}. Apart from this shift in the $q$ direction and a deformation of $\mathbb P$, the trajectories in the $\{q,\phi\}$ space are not qualitatively altered at this level of applied fields and all our previous considerations on the wall motion still hold. 


\subsection{Domain wall dynamics at different voltages }
So far we have modeled the DW motion at the sole current corresponding to the macrospin critical switching voltage $V_\mathrm{c}$. Changing the voltage will affect linearly the domain wall drift velocity $\bar{\Dot{q}}$ (Eq.~\ref{driftvelocity}) and thus change the transition time in an approximate linear manner, as indeed observed experimentally  \cite{hahn_time-resolved_2016, devolder_material_2018}. However it will also affect the oscillatory part of the domain wall motion by affecting the tilt dynamics (Eq.~\ref{omegaq}). Fig. S4 of the supplementary material shows how a change of the STT alters the trajectories in the $\{q,\phi\}$ space for a 80 nm disc. 

The most striking effects are a shift of the pond $\mathbb P$ in the $\phi$ direction, as well as a reduction of its overall size when increasing the magnitude of the STT. The shift in the $\phi$ direction can be understood from Eq.~\ref{model2b} because $\dot \phi$ has to vanish at the pond center. Therefore $\sigma j$ has to compensate the term $\frac{\partial U}{\partial \phi}$. To first order in STT, the $\phi$ shift is thus $\frac{\sigma j}{\gamma_0 H_{N \leftrightarrow B}}$.
As a result of the shrinking of $\mathbb P$, the proportion of DW trajectories with marked pauses before and after the crossing of the disc center should be reduced by increasing the spin-torque: the reproducibility of the wall propagation time is thus expected to improve with the magnitude of the STT. 
This is in line with the decrease of the complexity of the transition observed experimentally at applied voltages [Fig.~\ref{wavy100nm}](e-g)]. 

In the same line, very large spin-torque amplitudes i.e. with: 
\begin{equation}
\sigma j > \frac{1}{2}\gamma_0 H_{N \leftrightarrow B}    
\end{equation}
like in the example of [Fig.~S4(f)] entirely suppress the retention pond so that the wall passes the disc center ballistically whatever its tilt state. At these amplitudes of STT, the domain wall dynamics resembles that observed in 40 nm discs (Fig.~\ref{LLGS40nm}). With our material parameters, this always ballistic situation is encountered or all disc diameter in the 20-100 nm interval provided the applied spin-torque exceeds $\approx 3V_c$.

\section{Conclusion} \label{conclusion}
Our conclusions are summarized in Fig.~\ref{FigSummary}. 
We have studied how spin-torque induces the propagation of a wall across thin discs with perpendicular anisotropy. Micromagnetics were used to identify the wall motion scenarios for two representative disc diameters, 40 and 80 nm. The results were confronted with experiments for the largest devices, with a qualitative agreement. At small disc diameters, the wall sweeps across the device in an almost monotonous manner; the wall dynamics is essentially independent from the manner in which the magnetization is initialized. Conversely at larger diameters, the wall performs small range back-and-forth oscillations superimposed on the gradual drift. Depending on the initialization state of the wall, it crosses the disc center either in a "ballistic" direct manner or with variably marked pauses before and after the crossing of the center. Some specific initializing conditions of the wall can even result in the wall swinging across the disc center several times, which correlates with the growth of a strong non-uniformity of the wall tilt as well as a gyration of the overall magnetic texture. In experiments, these scenarios correspond respectively to ramp-like, featureless time-resolved conductance curves, or to curves with a pronounced oscillation of the conductance when at midway, or finally to conductance waveforms with multiple oscillations at midway.

We have then adapted a collective coordinate model in which the wall is described by its position $q$ and the magnetization tilt $\phi$ within the wall. We introduced the concept of the stretch field, whose elastic part (Eq.~\ref{stretchfield}) describes to the affinity of the wall to reduce its length, and whose demagnetizing part (Eq.~\ref{NBstretchfield}) describes the affinity of the wall to the reduce its dipolar energy by rotating its tilt, generally away from the N\'eel configuration. During the motion of the wall, part of the system's energy flows back and forth between the energy reservoirs (Eq.~\ref{UNBphi} and \ref{Uelastic}) associated to the two components of the stretch field; the wall velocity (Eq.~\ref{driftvelocity}, quantitative) and its oscillation (Eq.~\ref{omegaq}, indicative only) can be understood from this picture. 

In experiments, the transition time needed for the wall to sweep through the device varies stochastically.
This can be understood from the concept of the retention pond $\mathbb{P}$: a region in the $q-\phi$ space in which walls of proper tilt are transiently bound to the disc center. Walls having trajectories tangent to the pond make two pauses before and after crossing the disc center, thereby yielding switching times longer than average. The size of the retention pond (Eq.~\ref{pondwidth}) is correlated with the energy difference between Bloch and N\'eel walls when at the disc center. There exists a single "magic" disc diameter (Eq.~\ref{diameterForNoPond}), the DEVice OptimaL DiametER (DEVOLDER) for which the retention pond disappears. For this specific diameter, we predict that the wall shall cross the disc center in a ballistic manner independently of its tilt such that the time needed for a domain wall to sweep through the disc will get largely independent from its tilt \cite{bouquin_stochastic_2021}. This is expected to maximize the reproducibility of the wall dynamics, which is of great interest for magnetic memory applications.

\begin{acknowledgements}
This work was supported by IMEC's Industrial Affiliation Program on STT-MRAM devices. P.B., J.-V. K. and T.D. performed the analytical and numerical parts of this work. T.D., O.B. and P.B. performed the experiments. Samples were provided thanks to the work of O.B., S.R., G.K. and S.C..
\end{acknowledgements}


\begin{thebibliography}{35}%
\makeatletter
\providecommand \@ifxundefined [1]{%
 \@ifx{#1\undefined}
}%
\providecommand \@ifnum [1]{%
 \ifnum #1\expandafter \@firstoftwo
 \else \expandafter \@secondoftwo
 \fi
}%
\providecommand \@ifx [1]{%
 \ifx #1\expandafter \@firstoftwo
 \else \expandafter \@secondoftwo
 \fi
}%
\providecommand \natexlab [1]{#1}%
\providecommand \enquote  [1]{``#1''}%
\providecommand \bibnamefont  [1]{#1}%
\providecommand \bibfnamefont [1]{#1}%
\providecommand \citenamefont [1]{#1}%
\providecommand \href@noop [0]{\@secondoftwo}%
\providecommand \href [0]{\begingroup \@sanitize@url \@href}%
\providecommand \@href[1]{\@@startlink{#1}\@@href}%
\providecommand \@@href[1]{\endgroup#1\@@endlink}%
\providecommand \@sanitize@url [0]{\catcode `\\12\catcode `\$12\catcode
  `\&12\catcode `\#12\catcode `\^12\catcode `\_12\catcode `\%12\relax}%
\providecommand \@@startlink[1]{}%
\providecommand \@@endlink[0]{}%
\providecommand \url  [0]{\begingroup\@sanitize@url \@url }%
\providecommand \@url [1]{\endgroup\@href {#1}{\urlprefix }}%
\providecommand \urlprefix  [0]{URL }%
\providecommand \Eprint [0]{\href }%
\providecommand \doibase [0]{https://doi.org/}%
\providecommand \selectlanguage [0]{\@gobble}%
\providecommand \bibinfo  [0]{\@secondoftwo}%
\providecommand \bibfield  [0]{\@secondoftwo}%
\providecommand \translation [1]{[#1]}%
\providecommand \BibitemOpen [0]{}%
\providecommand \bibitemStop [0]{}%
\providecommand \bibitemNoStop [0]{.\EOS\space}%
\providecommand \EOS [0]{\spacefactor3000\relax}%
\providecommand \BibitemShut  [1]{\csname bibitem#1\endcsname}%
\let\auto@bib@innerbib\@empty
\bibitem [{\citenamefont {Novosad}\ \emph {et~al.}(2000)\citenamefont
  {Novosad}, \citenamefont {Otani}, \citenamefont {Ohsawa}, \citenamefont
  {Kim}, \citenamefont {Fukamichi}, \citenamefont {Koike}, \citenamefont
  {Maruyama}, \citenamefont {Kitakami},\ and\ \citenamefont
  {Shimada}}]{novosad_novel_2000}%
  \BibitemOpen
  \bibfield  {author} {\bibinfo {author} {\bibfnamefont {V.}~\bibnamefont
  {Novosad}}, \bibinfo {author} {\bibfnamefont {Y.}~\bibnamefont {Otani}},
  \bibinfo {author} {\bibfnamefont {A.}~\bibnamefont {Ohsawa}}, \bibinfo
  {author} {\bibfnamefont {S.~G.}\ \bibnamefont {Kim}}, \bibinfo {author}
  {\bibfnamefont {K.}~\bibnamefont {Fukamichi}}, \bibinfo {author}
  {\bibfnamefont {J.}~\bibnamefont {Koike}}, \bibinfo {author} {\bibfnamefont
  {K.}~\bibnamefont {Maruyama}}, \bibinfo {author} {\bibfnamefont
  {O.}~\bibnamefont {Kitakami}},\ and\ \bibinfo {author} {\bibfnamefont
  {Y.}~\bibnamefont {Shimada}},\ }\bibfield  {title} {\bibinfo {title} {Novel
  magnetostrictive memory device},\ }\href {https://doi.org/10.1063/1.372719}
  {\bibfield  {journal} {\bibinfo  {journal} {Journal of Applied Physics}\
  }\textbf {\bibinfo {volume} {87}},\ \bibinfo {pages} {6400} (\bibinfo {year}
  {2000})}\BibitemShut {NoStop}%
\bibitem [{\citenamefont {Ohno}\ \emph {et~al.}(2000)\citenamefont {Ohno},
  \citenamefont {Chiba}, \citenamefont {Matsukura}, \citenamefont {Omiya},
  \citenamefont {Abe}, \citenamefont {Dietl}, \citenamefont {Ohno},\ and\
  \citenamefont {Ohtani}}]{ohno_electric-field_2000}%
  \BibitemOpen
  \bibfield  {author} {\bibinfo {author} {\bibfnamefont {H.}~\bibnamefont
  {Ohno}}, \bibinfo {author} {\bibfnamefont {D.}~\bibnamefont {Chiba}},
  \bibinfo {author} {\bibfnamefont {F.}~\bibnamefont {Matsukura}}, \bibinfo
  {author} {\bibfnamefont {T.}~\bibnamefont {Omiya}}, \bibinfo {author}
  {\bibfnamefont {E.}~\bibnamefont {Abe}}, \bibinfo {author} {\bibfnamefont
  {T.}~\bibnamefont {Dietl}}, \bibinfo {author} {\bibfnamefont
  {Y.}~\bibnamefont {Ohno}},\ and\ \bibinfo {author} {\bibfnamefont
  {K.}~\bibnamefont {Ohtani}},\ }\bibfield  {title} {\bibinfo {title}
  {Electric-field control of ferromagnetism},\ }\href
  {https://doi.org/10.1038/35050040} {\bibfield  {journal} {\bibinfo  {journal}
  {Nature}\ }\textbf {\bibinfo {volume} {408}},\ \bibinfo {pages} {944}
  (\bibinfo {year} {2000})}\BibitemShut {NoStop}%
\bibitem [{\citenamefont {Beaurepaire}\ \emph {et~al.}(1996)\citenamefont
  {Beaurepaire}, \citenamefont {Merle}, \citenamefont {Daunois},\ and\
  \citenamefont {Bigot}}]{beaurepaire_ultrafast_1996}%
  \BibitemOpen
  \bibfield  {author} {\bibinfo {author} {\bibfnamefont {E.}~\bibnamefont
  {Beaurepaire}}, \bibinfo {author} {\bibfnamefont {J.-C.}\ \bibnamefont
  {Merle}}, \bibinfo {author} {\bibfnamefont {A.}~\bibnamefont {Daunois}},\
  and\ \bibinfo {author} {\bibfnamefont {J.-Y.}\ \bibnamefont {Bigot}},\
  }\bibfield  {title} {\bibinfo {title} {Ultrafast {Spin} {Dynamics} in
  {Ferromagnetic} {Nickel}},\ }\href
  {https://doi.org/10.1103/PhysRevLett.76.4250} {\bibfield  {journal} {\bibinfo
   {journal} {Physical Review Letters}\ }\textbf {\bibinfo {volume} {76}},\
  \bibinfo {pages} {4250} (\bibinfo {year} {1996})}\BibitemShut {NoStop}%
\bibitem [{\citenamefont {Chappert}\ \emph {et~al.}(2007)\citenamefont
  {Chappert}, \citenamefont {Fert},\ and\ \citenamefont
  {Dau}}]{chappert_emergence_2007}%
  \BibitemOpen
  \bibfield  {author} {\bibinfo {author} {\bibfnamefont {C.}~\bibnamefont
  {Chappert}}, \bibinfo {author} {\bibfnamefont {A.}~\bibnamefont {Fert}},\
  and\ \bibinfo {author} {\bibfnamefont {F.~N.~V.}\ \bibnamefont {Dau}},\
  }\bibfield  {title} {\bibinfo {title} {The emergence of spin electronics in
  data storage},\ }\href {https://doi.org/10.1038/nmat2024} {\bibfield
  {journal} {\bibinfo  {journal} {Nature Materials}\ }\textbf {\bibinfo
  {volume} {6}},\ \bibinfo {pages} {813} (\bibinfo {year} {2007})}\BibitemShut
  {NoStop}%
\bibitem [{\citenamefont {Khvalkovskiy}\ \emph {et~al.}(2013)\citenamefont
  {Khvalkovskiy}, \citenamefont {Apalkov}, \citenamefont {Watts}, \citenamefont
  {Chepulskii}, \citenamefont {Beach}, \citenamefont {Ong}, \citenamefont
  {Tang}, \citenamefont {Driskill-Smith}, \citenamefont {Butler}, \citenamefont
  {Visscher}, \citenamefont {Lottis}, \citenamefont {Chen}, \citenamefont
  {Nikitin},\ and\ \citenamefont {Krounbi}}]{khvalkovskiy_basic_2013}%
  \BibitemOpen
  \bibfield  {author} {\bibinfo {author} {\bibfnamefont {A.~V.}\ \bibnamefont
  {Khvalkovskiy}}, \bibinfo {author} {\bibfnamefont {D.}~\bibnamefont
  {Apalkov}}, \bibinfo {author} {\bibfnamefont {S.}~\bibnamefont {Watts}},
  \bibinfo {author} {\bibfnamefont {R.}~\bibnamefont {Chepulskii}}, \bibinfo
  {author} {\bibfnamefont {R.~S.}\ \bibnamefont {Beach}}, \bibinfo {author}
  {\bibfnamefont {A.}~\bibnamefont {Ong}}, \bibinfo {author} {\bibfnamefont
  {X.}~\bibnamefont {Tang}}, \bibinfo {author} {\bibfnamefont {A.}~\bibnamefont
  {Driskill-Smith}}, \bibinfo {author} {\bibfnamefont {W.~H.}\ \bibnamefont
  {Butler}}, \bibinfo {author} {\bibfnamefont {P.~B.}\ \bibnamefont
  {Visscher}}, \bibinfo {author} {\bibfnamefont {D.}~\bibnamefont {Lottis}},
  \bibinfo {author} {\bibfnamefont {E.}~\bibnamefont {Chen}}, \bibinfo {author}
  {\bibfnamefont {V.}~\bibnamefont {Nikitin}},\ and\ \bibinfo {author}
  {\bibfnamefont {M.}~\bibnamefont {Krounbi}},\ }\bibfield  {title} {\bibinfo
  {title} {Basic principles of {STT}-{MRAM} cell operation in memory arrays},\
  }\href {https://doi.org/10.1088/0022-3727/46/7/074001} {\bibfield  {journal}
  {\bibinfo  {journal} {Journal of Physics D: Applied Physics}\ }\textbf
  {\bibinfo {volume} {46}},\ \bibinfo {pages} {074001} (\bibinfo {year}
  {2013})},\ \bibinfo {note} {publisher: IOP Publishing}\BibitemShut {NoStop}%
\bibitem [{\citenamefont {Bouquin}\ \emph {et~al.}(2018)\citenamefont
  {Bouquin}, \citenamefont {Rao}, \citenamefont {Kar},\ and\ \citenamefont
  {Devolder}}]{bouquin_size_2018}%
  \BibitemOpen
  \bibfield  {author} {\bibinfo {author} {\bibfnamefont {P.}~\bibnamefont
  {Bouquin}}, \bibinfo {author} {\bibfnamefont {S.}~\bibnamefont {Rao}},
  \bibinfo {author} {\bibfnamefont {G.~S.}\ \bibnamefont {Kar}},\ and\ \bibinfo
  {author} {\bibfnamefont {T.}~\bibnamefont {Devolder}},\ }\bibfield  {title}
  {\bibinfo {title} {Size dependence of spin-torque switching in perpendicular
  magnetic tunnel junctions},\ }\href {https://doi.org/10.1063/1.5055741}
  {\bibfield  {journal} {\bibinfo  {journal} {Applied Physics Letters}\
  }\textbf {\bibinfo {volume} {113}},\ \bibinfo {pages} {222408} (\bibinfo
  {year} {2018})}\BibitemShut {NoStop}%
\bibitem [{\citenamefont {You}(2014)}]{you_switching_2014}%
  \BibitemOpen
  \bibfield  {author} {\bibinfo {author} {\bibfnamefont {C.-Y.}\ \bibnamefont
  {You}},\ }\bibfield  {title} {\bibinfo {title} {Switching current density
  reduction in perpendicular magnetic anisotropy spin transfer torque magnetic
  tunneling junctions},\ }\href {https://doi.org/10.1063/1.4862963} {\bibfield
  {journal} {\bibinfo  {journal} {Journal of Applied Physics}\ }\textbf
  {\bibinfo {volume} {115}},\ \bibinfo {pages} {043914} (\bibinfo {year}
  {2014})}\BibitemShut {NoStop}%
\bibitem [{\citenamefont {Chaves-O’Flynn}\ \emph {et~al.}(2015)\citenamefont
  {Chaves-O’Flynn}, \citenamefont {Wolf}, \citenamefont {Sun},\ and\
  \citenamefont {Kent}}]{chaves-oflynn_thermal_2015}%
  \BibitemOpen
  \bibfield  {author} {\bibinfo {author} {\bibfnamefont {G.~D.}\ \bibnamefont
  {Chaves-O’Flynn}}, \bibinfo {author} {\bibfnamefont {G.}~\bibnamefont
  {Wolf}}, \bibinfo {author} {\bibfnamefont {J.~Z.}\ \bibnamefont {Sun}},\ and\
  \bibinfo {author} {\bibfnamefont {A.~D.}\ \bibnamefont {Kent}},\ }\bibfield
  {title} {\bibinfo {title} {Thermal {Stability} of {Magnetic} {States} in
  {Circular} {Thin}-{Film} {Nanomagnets} with {Large} {Perpendicular}
  {Magnetic} {Anisotropy}},\ }\href
  {https://doi.org/10.1103/PhysRevApplied.4.024010} {\bibfield  {journal}
  {\bibinfo  {journal} {Physical Review Applied}\ }\textbf {\bibinfo {volume}
  {4}},\ \bibinfo {pages} {024010} (\bibinfo {year} {2015})}\BibitemShut
  {NoStop}%
\bibitem [{\citenamefont {Sun}(2000)}]{sun_spin-current_2000}%
  \BibitemOpen
  \bibfield  {author} {\bibinfo {author} {\bibfnamefont {J.~Z.}\ \bibnamefont
  {Sun}},\ }\bibfield  {title} {\bibinfo {title} {Spin-current interaction with
  a monodomain magnetic body: {A} model study},\ }\href
  {https://doi.org/10.1103/PhysRevB.62.570} {\bibfield  {journal} {\bibinfo
  {journal} {Physical Review B}\ }\textbf {\bibinfo {volume} {62}},\ \bibinfo
  {pages} {570} (\bibinfo {year} {2000})}\BibitemShut {NoStop}%
\bibitem [{\citenamefont {Butler}\ \emph {et~al.}(2012)\citenamefont {Butler},
  \citenamefont {Mewes}, \citenamefont {Mewes}, \citenamefont {Visscher},
  \citenamefont {Rippard}, \citenamefont {Russek},\ and\ \citenamefont
  {Heindl}}]{butler_switching_2012}%
  \BibitemOpen
  \bibfield  {author} {\bibinfo {author} {\bibfnamefont {W.~H.}\ \bibnamefont
  {Butler}}, \bibinfo {author} {\bibfnamefont {T.}~\bibnamefont {Mewes}},
  \bibinfo {author} {\bibfnamefont {C.~K.~A.}\ \bibnamefont {Mewes}}, \bibinfo
  {author} {\bibfnamefont {P.~B.}\ \bibnamefont {Visscher}}, \bibinfo {author}
  {\bibfnamefont {W.~H.}\ \bibnamefont {Rippard}}, \bibinfo {author}
  {\bibfnamefont {S.~E.}\ \bibnamefont {Russek}},\ and\ \bibinfo {author}
  {\bibfnamefont {R.}~\bibnamefont {Heindl}},\ }\bibfield  {title} {\bibinfo
  {title} {Switching {Distributions} for {Perpendicular} {Spin}-{Torque}
  {Devices} {Within} the {Macrospin} {Approximation}},\ }\href
  {https://doi.org/10.1109/TMAG.2012.2209122} {\bibfield  {journal} {\bibinfo
  {journal} {IEEE Transactions on Magnetics}\ }\textbf {\bibinfo {volume}
  {48}},\ \bibinfo {pages} {4684} (\bibinfo {year} {2012})}\BibitemShut
  {NoStop}%
\bibitem [{\citenamefont {Pinna}\ \emph {et~al.}(2013)\citenamefont {Pinna},
  \citenamefont {Kent},\ and\ \citenamefont
  {Stein}}]{pinna_spin-transfer_2013}%
  \BibitemOpen
  \bibfield  {author} {\bibinfo {author} {\bibfnamefont {D.}~\bibnamefont
  {Pinna}}, \bibinfo {author} {\bibfnamefont {A.~D.}\ \bibnamefont {Kent}},\
  and\ \bibinfo {author} {\bibfnamefont {D.~L.}\ \bibnamefont {Stein}},\
  }\bibfield  {title} {\bibinfo {title} {Spin-{Transfer} {Torque}
  {Magnetization} {Reversal} in {Uniaxial} {Nanomagnets} with {Thermal}
  {Noise}},\ }\href {https://doi.org/10.1063/1.4813488} {\bibfield  {journal}
  {\bibinfo  {journal} {Journal of Applied Physics}\ }\textbf {\bibinfo
  {volume} {114}},\ \bibinfo {pages} {033901} (\bibinfo {year} {2013})},\
  \bibinfo {note} {arXiv: 1210.7675}\BibitemShut {NoStop}%
\bibitem [{\citenamefont {Tomita}\ \emph {et~al.}(2013)\citenamefont {Tomita},
  \citenamefont {Miwa}, \citenamefont {Nozaki}, \citenamefont {Yamashita},
  \citenamefont {Nagase}, \citenamefont {Nishiyama}, \citenamefont {Kitagawa},
  \citenamefont {Yoshikawa}, \citenamefont {Daibou}, \citenamefont {Nagamine},
  \citenamefont {Kishi}, \citenamefont {Ikegawa}, \citenamefont {Shimomura},
  \citenamefont {Yoda},\ and\ \citenamefont {Suzuki}}]{tomita_unified_2013}%
  \BibitemOpen
  \bibfield  {author} {\bibinfo {author} {\bibfnamefont {H.}~\bibnamefont
  {Tomita}}, \bibinfo {author} {\bibfnamefont {S.}~\bibnamefont {Miwa}},
  \bibinfo {author} {\bibfnamefont {T.}~\bibnamefont {Nozaki}}, \bibinfo
  {author} {\bibfnamefont {S.}~\bibnamefont {Yamashita}}, \bibinfo {author}
  {\bibfnamefont {T.}~\bibnamefont {Nagase}}, \bibinfo {author} {\bibfnamefont
  {K.}~\bibnamefont {Nishiyama}}, \bibinfo {author} {\bibfnamefont
  {E.}~\bibnamefont {Kitagawa}}, \bibinfo {author} {\bibfnamefont
  {M.}~\bibnamefont {Yoshikawa}}, \bibinfo {author} {\bibfnamefont
  {T.}~\bibnamefont {Daibou}}, \bibinfo {author} {\bibfnamefont
  {M.}~\bibnamefont {Nagamine}}, \bibinfo {author} {\bibfnamefont
  {T.}~\bibnamefont {Kishi}}, \bibinfo {author} {\bibfnamefont
  {S.}~\bibnamefont {Ikegawa}}, \bibinfo {author} {\bibfnamefont
  {N.}~\bibnamefont {Shimomura}}, \bibinfo {author} {\bibfnamefont
  {H.}~\bibnamefont {Yoda}},\ and\ \bibinfo {author} {\bibfnamefont
  {Y.}~\bibnamefont {Suzuki}},\ }\bibfield  {title} {\bibinfo {title} {Unified
  understanding of both thermally assisted and precessional spin-transfer
  switching in perpendicularly magnetized giant magnetoresistive nanopillars},\
  }\href {https://doi.org/10.1063/1.4789879} {\bibfield  {journal} {\bibinfo
  {journal} {Applied Physics Letters}\ }\textbf {\bibinfo {volume} {102}},\
  \bibinfo {pages} {042409} (\bibinfo {year} {2013})}\BibitemShut {NoStop}%
\bibitem [{\citenamefont {Bernstein}\ \emph {et~al.}(2011)\citenamefont
  {Bernstein}, \citenamefont {Bräuer}, \citenamefont {Kukreja}, \citenamefont
  {Stöhr}, \citenamefont {Hauet}, \citenamefont {Cucchiara}, \citenamefont
  {Mangin}, \citenamefont {Katine}, \citenamefont {Tyliszczak}, \citenamefont
  {Chou},\ and\ \citenamefont {Acremann}}]{bernstein_nonuniform_2011}%
  \BibitemOpen
  \bibfield  {author} {\bibinfo {author} {\bibfnamefont {D.~P.}\ \bibnamefont
  {Bernstein}}, \bibinfo {author} {\bibfnamefont {B.}~\bibnamefont {Bräuer}},
  \bibinfo {author} {\bibfnamefont {R.}~\bibnamefont {Kukreja}}, \bibinfo
  {author} {\bibfnamefont {J.}~\bibnamefont {Stöhr}}, \bibinfo {author}
  {\bibfnamefont {T.}~\bibnamefont {Hauet}}, \bibinfo {author} {\bibfnamefont
  {J.}~\bibnamefont {Cucchiara}}, \bibinfo {author} {\bibfnamefont
  {S.}~\bibnamefont {Mangin}}, \bibinfo {author} {\bibfnamefont {J.~A.}\
  \bibnamefont {Katine}}, \bibinfo {author} {\bibfnamefont {T.}~\bibnamefont
  {Tyliszczak}}, \bibinfo {author} {\bibfnamefont {K.~W.}\ \bibnamefont
  {Chou}},\ and\ \bibinfo {author} {\bibfnamefont {Y.}~\bibnamefont
  {Acremann}},\ }\bibfield  {title} {\bibinfo {title} {Nonuniform switching of
  the perpendicular magnetization in a spin-torque-driven magnetic
  nanopillar},\ }\href {https://doi.org/10.1103/PhysRevB.83.180410} {\bibfield
  {journal} {\bibinfo  {journal} {Physical Review B}\ }\textbf {\bibinfo
  {volume} {83}},\ \bibinfo {pages} {180410} (\bibinfo {year}
  {2011})}\BibitemShut {NoStop}%
\bibitem [{\citenamefont {Sun}\ \emph {et~al.}(2013)\citenamefont {Sun},
  \citenamefont {Brown}, \citenamefont {Chen}, \citenamefont {Delenia},
  \citenamefont {Gaidis}, \citenamefont {Harms}, \citenamefont {Hu},
  \citenamefont {Jiang}, \citenamefont {Kilaru}, \citenamefont {Kula},
  \citenamefont {Lauer}, \citenamefont {Liu}, \citenamefont {Murthy},
  \citenamefont {Nowak}, \citenamefont {O’Sullivan}, \citenamefont {Parkin},
  \citenamefont {Robertazzi}, \citenamefont {Rice}, \citenamefont {Sandhu},
  \citenamefont {Topuria},\ and\ \citenamefont
  {Worledge}}]{sun_spin-torque_2013}%
  \BibitemOpen
  \bibfield  {author} {\bibinfo {author} {\bibfnamefont {J.~Z.}\ \bibnamefont
  {Sun}}, \bibinfo {author} {\bibfnamefont {S.~L.}\ \bibnamefont {Brown}},
  \bibinfo {author} {\bibfnamefont {W.}~\bibnamefont {Chen}}, \bibinfo {author}
  {\bibfnamefont {E.~A.}\ \bibnamefont {Delenia}}, \bibinfo {author}
  {\bibfnamefont {M.~C.}\ \bibnamefont {Gaidis}}, \bibinfo {author}
  {\bibfnamefont {J.}~\bibnamefont {Harms}}, \bibinfo {author} {\bibfnamefont
  {G.}~\bibnamefont {Hu}}, \bibinfo {author} {\bibfnamefont {X.}~\bibnamefont
  {Jiang}}, \bibinfo {author} {\bibfnamefont {R.}~\bibnamefont {Kilaru}},
  \bibinfo {author} {\bibfnamefont {W.}~\bibnamefont {Kula}}, \bibinfo {author}
  {\bibfnamefont {G.}~\bibnamefont {Lauer}}, \bibinfo {author} {\bibfnamefont
  {L.~Q.}\ \bibnamefont {Liu}}, \bibinfo {author} {\bibfnamefont
  {S.}~\bibnamefont {Murthy}}, \bibinfo {author} {\bibfnamefont
  {J.}~\bibnamefont {Nowak}}, \bibinfo {author} {\bibfnamefont {E.~J.}\
  \bibnamefont {O’Sullivan}}, \bibinfo {author} {\bibfnamefont {S.~S.~P.}\
  \bibnamefont {Parkin}}, \bibinfo {author} {\bibfnamefont {R.~P.}\
  \bibnamefont {Robertazzi}}, \bibinfo {author} {\bibfnamefont {P.~M.}\
  \bibnamefont {Rice}}, \bibinfo {author} {\bibfnamefont {G.}~\bibnamefont
  {Sandhu}}, \bibinfo {author} {\bibfnamefont {T.}~\bibnamefont {Topuria}},\
  and\ \bibinfo {author} {\bibfnamefont {D.~C.}\ \bibnamefont {Worledge}},\
  }\bibfield  {title} {\bibinfo {title} {Spin-torque switching efficiency in
  {CoFeB}-{MgO} based tunnel junctions},\ }\href
  {https://doi.org/10.1103/PhysRevB.88.104426} {\bibfield  {journal} {\bibinfo
  {journal} {Physical Review B}\ }\textbf {\bibinfo {volume} {88}},\ \bibinfo
  {pages} {104426} (\bibinfo {year} {2013})}\BibitemShut {NoStop}%
\bibitem [{\citenamefont {Hahn}\ \emph {et~al.}(2016)\citenamefont {Hahn},
  \citenamefont {Wolf}, \citenamefont {Kardasz}, \citenamefont {Watts},
  \citenamefont {Pinarbasi},\ and\ \citenamefont
  {Kent}}]{hahn_time-resolved_2016}%
  \BibitemOpen
  \bibfield  {author} {\bibinfo {author} {\bibfnamefont {C.}~\bibnamefont
  {Hahn}}, \bibinfo {author} {\bibfnamefont {G.}~\bibnamefont {Wolf}}, \bibinfo
  {author} {\bibfnamefont {B.}~\bibnamefont {Kardasz}}, \bibinfo {author}
  {\bibfnamefont {S.}~\bibnamefont {Watts}}, \bibinfo {author} {\bibfnamefont
  {M.}~\bibnamefont {Pinarbasi}},\ and\ \bibinfo {author} {\bibfnamefont
  {A.~D.}\ \bibnamefont {Kent}},\ }\bibfield  {title} {\bibinfo {title}
  {Time-resolved studies of the spin-transfer reversal mechanism in
  perpendicularly magnetized magnetic tunnel junctions},\ }\href
  {https://doi.org/10.1103/PhysRevB.94.214432} {\bibfield  {journal} {\bibinfo
  {journal} {Physical Review B}\ }\textbf {\bibinfo {volume} {94}},\ \bibinfo
  {pages} {214432} (\bibinfo {year} {2016})}\BibitemShut {NoStop}%
\bibitem [{\citenamefont {Pizzini}\ \emph {et~al.}(2014)\citenamefont
  {Pizzini}, \citenamefont {Vogel}, \citenamefont {Rohart}, \citenamefont
  {Buda-Prejbeanu}, \citenamefont {Jué}, \citenamefont {Boulle}, \citenamefont
  {Miron}, \citenamefont {Safeer}, \citenamefont {Auffret}, \citenamefont
  {Gaudin},\ and\ \citenamefont {Thiaville}}]{pizzini_chirality-induced_2014}%
  \BibitemOpen
  \bibfield  {author} {\bibinfo {author} {\bibfnamefont {S.}~\bibnamefont
  {Pizzini}}, \bibinfo {author} {\bibfnamefont {J.}~\bibnamefont {Vogel}},
  \bibinfo {author} {\bibfnamefont {S.}~\bibnamefont {Rohart}}, \bibinfo
  {author} {\bibfnamefont {L.}~\bibnamefont {Buda-Prejbeanu}}, \bibinfo
  {author} {\bibfnamefont {E.}~\bibnamefont {Jué}}, \bibinfo {author}
  {\bibfnamefont {O.}~\bibnamefont {Boulle}}, \bibinfo {author} {\bibfnamefont
  {I.}~\bibnamefont {Miron}}, \bibinfo {author} {\bibfnamefont
  {C.}~\bibnamefont {Safeer}}, \bibinfo {author} {\bibfnamefont
  {S.}~\bibnamefont {Auffret}}, \bibinfo {author} {\bibfnamefont
  {G.}~\bibnamefont {Gaudin}},\ and\ \bibinfo {author} {\bibfnamefont
  {A.}~\bibnamefont {Thiaville}},\ }\bibfield  {title} {\bibinfo {title}
  {Chirality-{Induced} {Asymmetric} {Magnetic} {Nucleation} in {Pt}/{Co}/{AlOx}
  {Ultrathin} {Microstructures}},\ }\href
  {https://doi.org/10.1103/PhysRevLett.113.047203} {\bibfield  {journal}
  {\bibinfo  {journal} {Physical Review Letters}\ }\textbf {\bibinfo {volume}
  {113}},\ \bibinfo {pages} {047203} (\bibinfo {year} {2014})},\ \bibinfo
  {note} {publisher: American Physical Society}\BibitemShut {NoStop}%
\bibitem [{\citenamefont {Munira}\ and\ \citenamefont
  {Visscher}(2015)}]{munira_calculation_2015}%
  \BibitemOpen
  \bibfield  {author} {\bibinfo {author} {\bibfnamefont {K.}~\bibnamefont
  {Munira}}\ and\ \bibinfo {author} {\bibfnamefont {P.~B.}\ \bibnamefont
  {Visscher}},\ }\bibfield  {title} {\bibinfo {title} {Calculation of
  energy-barrier lowering by incoherent switching in spin-transfer torque
  magnetoresistive random-access memory},\ }\href
  {https://doi.org/10.1063/1.4908153} {\bibfield  {journal} {\bibinfo
  {journal} {Journal of Applied Physics}\ }\textbf {\bibinfo {volume} {117}},\
  \bibinfo {pages} {17B710} (\bibinfo {year} {2015})}\BibitemShut {NoStop}%
\bibitem [{\citenamefont {Visscher}\ \emph {et~al.}(2016)\citenamefont
  {Visscher}, \citenamefont {Munira},\ and\ \citenamefont
  {Rosati}}]{visscher_instability_2016}%
  \BibitemOpen
  \bibfield  {author} {\bibinfo {author} {\bibfnamefont {P.~B.}\ \bibnamefont
  {Visscher}}, \bibinfo {author} {\bibfnamefont {K.}~\bibnamefont {Munira}},\
  and\ \bibinfo {author} {\bibfnamefont {R.~J.}\ \bibnamefont {Rosati}},\
  }\bibfield  {title} {\bibinfo {title} {Instability {Mechanism} for
  {STT}-{MRAM} switching},\ }\href {http://arxiv.org/abs/1604.03992} {\bibfield
   {journal} {\bibinfo  {journal} {arXiv:1604.03992 [cond-mat]}\ } (\bibinfo
  {year} {2016})},\ \bibinfo {note} {arXiv: 1604.03992}\BibitemShut {NoStop}%
\bibitem [{\citenamefont {Devolder}\ \emph
  {et~al.}(2016{\natexlab{a}})\citenamefont {Devolder}, \citenamefont {Kim},
  \citenamefont {Garcia-Sanchez}, \citenamefont {Swerts}, \citenamefont {Kim},
  \citenamefont {Couet}, \citenamefont {Kar},\ and\ \citenamefont
  {Furnemont}}]{devolder_time-resolved_2016}%
  \BibitemOpen
  \bibfield  {author} {\bibinfo {author} {\bibfnamefont {T.}~\bibnamefont
  {Devolder}}, \bibinfo {author} {\bibfnamefont {J.-V.}\ \bibnamefont {Kim}},
  \bibinfo {author} {\bibfnamefont {F.}~\bibnamefont {Garcia-Sanchez}},
  \bibinfo {author} {\bibfnamefont {J.}~\bibnamefont {Swerts}}, \bibinfo
  {author} {\bibfnamefont {W.}~\bibnamefont {Kim}}, \bibinfo {author}
  {\bibfnamefont {S.}~\bibnamefont {Couet}}, \bibinfo {author} {\bibfnamefont
  {G.}~\bibnamefont {Kar}},\ and\ \bibinfo {author} {\bibfnamefont
  {A.}~\bibnamefont {Furnemont}},\ }\bibfield  {title} {\bibinfo {title}
  {Time-resolved spin-torque switching in {MgO}-based perpendicularly
  magnetized tunnel junctions},\ }\href
  {https://doi.org/10.1103/PhysRevB.93.024420} {\bibfield  {journal} {\bibinfo
  {journal} {Physical Review B}\ }\textbf {\bibinfo {volume} {93}},\ \bibinfo
  {pages} {024420} (\bibinfo {year} {2016}{\natexlab{a}})}\BibitemShut
  {NoStop}%
\bibitem [{\citenamefont {Tomita}\ \emph {et~al.}(2011)\citenamefont {Tomita},
  \citenamefont {Nozaki}, \citenamefont {Seki}, \citenamefont {Nagase},
  \citenamefont {Nishiyama}, \citenamefont {Kitagawa}, \citenamefont
  {Yoshikawa}, \citenamefont {Daibou}, \citenamefont {Nagamine},\ and\
  \citenamefont {Kishi}}]{tomita_high-speed_2011}%
  \BibitemOpen
  \bibfield  {author} {\bibinfo {author} {\bibfnamefont {H.}~\bibnamefont
  {Tomita}}, \bibinfo {author} {\bibfnamefont {T.}~\bibnamefont {Nozaki}},
  \bibinfo {author} {\bibfnamefont {T.}~\bibnamefont {Seki}}, \bibinfo {author}
  {\bibfnamefont {T.}~\bibnamefont {Nagase}}, \bibinfo {author} {\bibfnamefont
  {K.}~\bibnamefont {Nishiyama}}, \bibinfo {author} {\bibfnamefont
  {E.}~\bibnamefont {Kitagawa}}, \bibinfo {author} {\bibfnamefont
  {M.}~\bibnamefont {Yoshikawa}}, \bibinfo {author} {\bibfnamefont
  {T.}~\bibnamefont {Daibou}}, \bibinfo {author} {\bibfnamefont
  {M.}~\bibnamefont {Nagamine}},\ and\ \bibinfo {author} {\bibfnamefont
  {T.}~\bibnamefont {Kishi}},\ }\bibfield  {title} {\bibinfo {title}
  {High-speed spin-transfer switching in {GMR} nano-pillars with perpendicular
  anisotropy},\ }\href@noop {} {\bibfield  {journal} {\bibinfo  {journal} {IEEE
  Transactions on Magnetics}\ }\textbf {\bibinfo {volume} {47}},\ \bibinfo
  {pages} {1599} (\bibinfo {year} {2011})}\BibitemShut {NoStop}%
\bibitem [{\citenamefont {Devolder}\ \emph {et~al.}(2018)\citenamefont
  {Devolder}, \citenamefont {Kim}, \citenamefont {Swerts}, \citenamefont
  {Couet}, \citenamefont {Rao}, \citenamefont {Kim}, \citenamefont {Mertens},
  \citenamefont {Kar},\ and\ \citenamefont {Nikitin}}]{devolder_material_2018}%
  \BibitemOpen
  \bibfield  {author} {\bibinfo {author} {\bibfnamefont {T.}~\bibnamefont
  {Devolder}}, \bibinfo {author} {\bibfnamefont {J.~V.}\ \bibnamefont {Kim}},
  \bibinfo {author} {\bibfnamefont {J.}~\bibnamefont {Swerts}}, \bibinfo
  {author} {\bibfnamefont {S.}~\bibnamefont {Couet}}, \bibinfo {author}
  {\bibfnamefont {S.}~\bibnamefont {Rao}}, \bibinfo {author} {\bibfnamefont
  {W.}~\bibnamefont {Kim}}, \bibinfo {author} {\bibfnamefont {S.}~\bibnamefont
  {Mertens}}, \bibinfo {author} {\bibfnamefont {G.}~\bibnamefont {Kar}},\ and\
  \bibinfo {author} {\bibfnamefont {V.}~\bibnamefont {Nikitin}},\ }\bibfield
  {title} {\bibinfo {title} {Material {Developments} and {Domain}
  {Wall}-{Based} {Nanosecond}-{Scale} {Switching} {Process} in
  {Perpendicularly} {Magnetized} {STT}-{MRAM} {Cells}},\ }\href
  {https://doi.org/10.1109/TMAG.2017.2739187} {\bibfield  {journal} {\bibinfo
  {journal} {IEEE Transactions on Magnetics}\ }\textbf {\bibinfo {volume}
  {54}},\ \bibinfo {pages} {1} (\bibinfo {year} {2018})}\BibitemShut {NoStop}%
\bibitem [{\citenamefont {Devolder}\ \emph
  {et~al.}(2016{\natexlab{b}})\citenamefont {Devolder}, \citenamefont
  {Le~Goff},\ and\ \citenamefont {Nikitin}}]{devolder_size_2016}%
  \BibitemOpen
  \bibfield  {author} {\bibinfo {author} {\bibfnamefont {T.}~\bibnamefont
  {Devolder}}, \bibinfo {author} {\bibfnamefont {A.}~\bibnamefont {Le~Goff}},\
  and\ \bibinfo {author} {\bibfnamefont {V.}~\bibnamefont {Nikitin}},\
  }\bibfield  {title} {\bibinfo {title} {Size dependence of nanosecond-scale
  spin-torque switching in perpendicularly magnetized tunnel junctions},\
  }\href {https://doi.org/10.1103/PhysRevB.93.224432} {\bibfield  {journal}
  {\bibinfo  {journal} {Physical Review B}\ }\textbf {\bibinfo {volume} {93}},\
  \bibinfo {pages} {224432} (\bibinfo {year} {2016}{\natexlab{b}})}\BibitemShut
  {NoStop}%
\bibitem [{\citenamefont {Bouquin}\ \emph {et~al.}(2021)\citenamefont
  {Bouquin}, \citenamefont {Kim}, \citenamefont {Bultynck}, \citenamefont
  {Rao}, \citenamefont {Couet}, \citenamefont {Kar},\ and\ \citenamefont
  {Devolder}}]{bouquin_stochastic_2021}%
  \BibitemOpen
  \bibfield  {author} {\bibinfo {author} {\bibfnamefont {P.}~\bibnamefont
  {Bouquin}}, \bibinfo {author} {\bibfnamefont {J.-V.}\ \bibnamefont {Kim}},
  \bibinfo {author} {\bibfnamefont {O.}~\bibnamefont {Bultynck}}, \bibinfo
  {author} {\bibfnamefont {S.}~\bibnamefont {Rao}}, \bibinfo {author}
  {\bibfnamefont {S.}~\bibnamefont {Couet}}, \bibinfo {author} {\bibfnamefont
  {G.~S.}\ \bibnamefont {Kar}},\ and\ \bibinfo {author} {\bibfnamefont
  {T.}~\bibnamefont {Devolder}},\ }\bibfield  {title} {\bibinfo {title}
  {Stochastic {Processes} in {Magnetization} {Reversal} {Involving}
  {Domain}-{Wall} {Motion} in {Magnetic} {Memory} {Elements}},\ }\href
  {https://doi.org/10.1103/PhysRevApplied.15.024037} {\bibfield  {journal}
  {\bibinfo  {journal} {Physical Review Applied}\ }\textbf {\bibinfo {volume}
  {15}},\ \bibinfo {pages} {024037} (\bibinfo {year} {2021})}\BibitemShut
  {NoStop}%
\bibitem [{\citenamefont {Statuto}\ \emph {et~al.}(2021)\citenamefont
  {Statuto}, \citenamefont {Mohammadi},\ and\ \citenamefont
  {Kent}}]{statuto_micromagnetic_2021}%
  \BibitemOpen
  \bibfield  {author} {\bibinfo {author} {\bibfnamefont {N.}~\bibnamefont
  {Statuto}}, \bibinfo {author} {\bibfnamefont {J.~B.}\ \bibnamefont
  {Mohammadi}},\ and\ \bibinfo {author} {\bibfnamefont {A.~D.}\ \bibnamefont
  {Kent}},\ }\bibfield  {title} {\bibinfo {title} {Micromagnetic instabilities
  in spin-transfer switching of perpendicular magnetic tunnel junctions},\
  }\href {https://doi.org/10.1103/PhysRevB.103.014409} {\bibfield  {journal}
  {\bibinfo  {journal} {Physical Review B}\ }\textbf {\bibinfo {volume}
  {103}},\ \bibinfo {pages} {014409} (\bibinfo {year} {2021})}\BibitemShut
  {NoStop}%
\bibitem [{\citenamefont {Devolder}\ \emph {et~al.}(2019)\citenamefont
  {Devolder}, \citenamefont {Carpenter}, \citenamefont {Rao}, \citenamefont
  {Kim}, \citenamefont {Couet}, \citenamefont {Swerts},\ and\ \citenamefont
  {Kar}}]{devolder_offset_2019}%
  \BibitemOpen
  \bibfield  {author} {\bibinfo {author} {\bibfnamefont {T.}~\bibnamefont
  {Devolder}}, \bibinfo {author} {\bibfnamefont {R.}~\bibnamefont {Carpenter}},
  \bibinfo {author} {\bibfnamefont {S.}~\bibnamefont {Rao}}, \bibinfo {author}
  {\bibfnamefont {W.}~\bibnamefont {Kim}}, \bibinfo {author} {\bibfnamefont
  {S.}~\bibnamefont {Couet}}, \bibinfo {author} {\bibfnamefont
  {J.}~\bibnamefont {Swerts}},\ and\ \bibinfo {author} {\bibfnamefont {G.~S.}\
  \bibnamefont {Kar}},\ }\bibfield  {title} {\bibinfo {title} {Offset fields in
  perpendicularly magnetized tunnel junctions},\ }\href
  {https://doi.org/10.1088%2F1361-6463%2Fab1b07} {\bibfield  {journal}
  {\bibinfo  {journal} {Journal of Physics D: Applied Physics}\ }\textbf
  {\bibinfo {volume} {52}},\ \bibinfo {pages} {274001} (\bibinfo {year}
  {2019})}\BibitemShut {NoStop}%
\bibitem [{\citenamefont {Devolder}\ \emph {et~al.}(2008)\citenamefont
  {Devolder}, \citenamefont {Hayakawa}, \citenamefont {Ito}, \citenamefont
  {Takahashi}, \citenamefont {Ikeda}, \citenamefont {Crozat}, \citenamefont
  {Zerounian}, \citenamefont {Kim}, \citenamefont {Chappert},\ and\
  \citenamefont {Ohno}}]{devolder_single-shot_2008}%
  \BibitemOpen
  \bibfield  {author} {\bibinfo {author} {\bibfnamefont {T.}~\bibnamefont
  {Devolder}}, \bibinfo {author} {\bibfnamefont {J.}~\bibnamefont {Hayakawa}},
  \bibinfo {author} {\bibfnamefont {K.}~\bibnamefont {Ito}}, \bibinfo {author}
  {\bibfnamefont {H.}~\bibnamefont {Takahashi}}, \bibinfo {author}
  {\bibfnamefont {S.}~\bibnamefont {Ikeda}}, \bibinfo {author} {\bibfnamefont
  {P.}~\bibnamefont {Crozat}}, \bibinfo {author} {\bibfnamefont
  {N.}~\bibnamefont {Zerounian}}, \bibinfo {author} {\bibfnamefont {J.-V.}\
  \bibnamefont {Kim}}, \bibinfo {author} {\bibfnamefont {C.}~\bibnamefont
  {Chappert}},\ and\ \bibinfo {author} {\bibfnamefont {H.}~\bibnamefont
  {Ohno}},\ }\bibfield  {title} {\bibinfo {title} {Single-{Shot}
  {Time}-{Resolved} {Measurements} of {Nanosecond}-{Scale} {Spin}-{Transfer}
  {Induced} {Switching}: {Stochastic} {Versus} {Deterministic} {Aspects}},\
  }\href {https://doi.org/10.1103/PhysRevLett.100.057206} {\bibfield  {journal}
  {\bibinfo  {journal} {Physical Review Letters}\ }\textbf {\bibinfo {volume}
  {100}},\ \bibinfo {pages} {057206} (\bibinfo {year} {2008})}\BibitemShut
  {NoStop}%
\bibitem [{\citenamefont {Bultynck}\ \emph {et~al.}(2018)\citenamefont
  {Bultynck}, \citenamefont {Manfrini}, \citenamefont {Vaysset}, \citenamefont
  {Swerts}, \citenamefont {Wilson}, \citenamefont {Sorée}, \citenamefont
  {Heyns}, \citenamefont {Mocuta}, \citenamefont {Radu},\ and\ \citenamefont
  {Devolder}}]{bultynck_instant-spin_2018}%
  \BibitemOpen
  \bibfield  {author} {\bibinfo {author} {\bibfnamefont {O.}~\bibnamefont
  {Bultynck}}, \bibinfo {author} {\bibfnamefont {M.}~\bibnamefont {Manfrini}},
  \bibinfo {author} {\bibfnamefont {A.}~\bibnamefont {Vaysset}}, \bibinfo
  {author} {\bibfnamefont {J.}~\bibnamefont {Swerts}}, \bibinfo {author}
  {\bibfnamefont {C.~J.}\ \bibnamefont {Wilson}}, \bibinfo {author}
  {\bibfnamefont {B.}~\bibnamefont {Sorée}}, \bibinfo {author} {\bibfnamefont
  {M.}~\bibnamefont {Heyns}}, \bibinfo {author} {\bibfnamefont
  {D.}~\bibnamefont {Mocuta}}, \bibinfo {author} {\bibfnamefont {I.~P.}\
  \bibnamefont {Radu}},\ and\ \bibinfo {author} {\bibfnamefont
  {T.}~\bibnamefont {Devolder}},\ }\bibfield  {title} {\bibinfo {title}
  {Instant-{On} {Spin} {Torque} in {Noncollinear} {Magnetic} {Tunnel}
  {Junctions}},\ }\href {https://doi.org/10.1103/PhysRevApplied.10.054028}
  {\bibfield  {journal} {\bibinfo  {journal} {Physical Review Applied}\
  }\textbf {\bibinfo {volume} {10}},\ \bibinfo {pages} {054028} (\bibinfo
  {year} {2018})}\BibitemShut {NoStop}%
\bibitem [{\citenamefont {Krizakova}\ \emph {et~al.}(2019)\citenamefont
  {Krizakova}, \citenamefont {Garcia}, \citenamefont {Vogel}, \citenamefont
  {Rougemaille}, \citenamefont {Chaves}, \citenamefont {Pizzini},\ and\
  \citenamefont {Thiaville}}]{krizakova_study_2019}%
  \BibitemOpen
  \bibfield  {author} {\bibinfo {author} {\bibfnamefont {V.}~\bibnamefont
  {Krizakova}}, \bibinfo {author} {\bibfnamefont {J.~P.}\ \bibnamefont
  {Garcia}}, \bibinfo {author} {\bibfnamefont {J.}~\bibnamefont {Vogel}},
  \bibinfo {author} {\bibfnamefont {N.}~\bibnamefont {Rougemaille}}, \bibinfo
  {author} {\bibfnamefont {D.~d.~S.}\ \bibnamefont {Chaves}}, \bibinfo {author}
  {\bibfnamefont {S.}~\bibnamefont {Pizzini}},\ and\ \bibinfo {author}
  {\bibfnamefont {A.}~\bibnamefont {Thiaville}},\ }\bibfield  {title} {\bibinfo
  {title} {Study of the velocity plateau of {Dzyaloshinskii} domain walls},\
  }\href {https://doi.org/10.1103/PhysRevB.100.214404} {\bibfield  {journal}
  {\bibinfo  {journal} {Physical Review B}\ }\textbf {\bibinfo {volume}
  {100}},\ \bibinfo {pages} {214404} (\bibinfo {year} {2019})}\BibitemShut
  {NoStop}%
\bibitem [{\citenamefont {Vansteenkiste}\ \emph {et~al.}(2014)\citenamefont
  {Vansteenkiste}, \citenamefont {Leliaert}, \citenamefont {Dvornik},
  \citenamefont {Helsen}, \citenamefont {Garcia-Sanchez},\ and\ \citenamefont
  {Waeyenberge}}]{vansteenkiste_design_2014}%
  \BibitemOpen
  \bibfield  {author} {\bibinfo {author} {\bibfnamefont {A.}~\bibnamefont
  {Vansteenkiste}}, \bibinfo {author} {\bibfnamefont {J.}~\bibnamefont
  {Leliaert}}, \bibinfo {author} {\bibfnamefont {M.}~\bibnamefont {Dvornik}},
  \bibinfo {author} {\bibfnamefont {M.}~\bibnamefont {Helsen}}, \bibinfo
  {author} {\bibfnamefont {F.}~\bibnamefont {Garcia-Sanchez}},\ and\ \bibinfo
  {author} {\bibfnamefont {B.~V.}\ \bibnamefont {Waeyenberge}},\ }\bibfield
  {title} {\bibinfo {title} {The design and verification of {MuMax3}},\ }\href
  {https://doi.org/10.1063/1.4899186} {\bibfield  {journal} {\bibinfo
  {journal} {AIP Advances}\ }\textbf {\bibinfo {volume} {4}},\ \bibinfo {pages}
  {107133} (\bibinfo {year} {2014})}\BibitemShut {NoStop}%
\bibitem [{\citenamefont {Thiaville}\ \emph {et~al.}(2002)\citenamefont
  {Thiaville}, \citenamefont {Garcia},\ and\ \citenamefont
  {Miltat}}]{thiaville_domain_2002}%
  \BibitemOpen
  \bibfield  {author} {\bibinfo {author} {\bibfnamefont {A.}~\bibnamefont
  {Thiaville}}, \bibinfo {author} {\bibfnamefont {J.~M.}\ \bibnamefont
  {Garcia}},\ and\ \bibinfo {author} {\bibfnamefont {J.}~\bibnamefont
  {Miltat}},\ }\bibfield  {title} {\bibinfo {title} {Domain wall dynamics in
  nanowires},\ }\href {https://doi.org/10.1016/S0304-8853(01)01353-1}
  {\bibfield  {journal} {\bibinfo  {journal} {Journal of Magnetism and Magnetic
  Materials}\ }\bibinfo {series} {Proceedings of the {Joint} {European}
  {Magnetic} {Symposia} ({JEMS01})},\ \textbf {\bibinfo {volume} {242-245}},\
  \bibinfo {pages} {1061} (\bibinfo {year} {2002})}\BibitemShut {NoStop}%
\bibitem [{\citenamefont {Thiaville}\ \emph {et~al.}(2004)\citenamefont
  {Thiaville}, \citenamefont {Nakatani}, \citenamefont {Miltat},\ and\
  \citenamefont {Vernier}}]{thiaville_domain_2004}%
  \BibitemOpen
  \bibfield  {author} {\bibinfo {author} {\bibfnamefont {A.}~\bibnamefont
  {Thiaville}}, \bibinfo {author} {\bibfnamefont {Y.}~\bibnamefont {Nakatani}},
  \bibinfo {author} {\bibfnamefont {J.}~\bibnamefont {Miltat}},\ and\ \bibinfo
  {author} {\bibfnamefont {N.}~\bibnamefont {Vernier}},\ }\bibfield  {title}
  {\bibinfo {title} {Domain wall motion by spin-polarized current: a
  micromagnetic study},\ }\href {https://doi.org/10.1063/1.1667804} {\bibfield
  {journal} {\bibinfo  {journal} {Journal of Applied Physics}\ }\textbf
  {\bibinfo {volume} {95}},\ \bibinfo {pages} {7049} (\bibinfo {year}
  {2004})}\BibitemShut {NoStop}%
\bibitem [{\citenamefont {Cucchiara}\ \emph {et~al.}(2012)\citenamefont
  {Cucchiara}, \citenamefont {Le~Gall}, \citenamefont {Fullerton},
  \citenamefont {Kim}, \citenamefont {Ravelosona}, \citenamefont {Henry},
  \citenamefont {Katine}, \citenamefont {Kent}, \citenamefont {Bedau},
  \citenamefont {Gopman},\ and\ \citenamefont
  {Mangin}}]{cucchiara_domain_2012}%
  \BibitemOpen
  \bibfield  {author} {\bibinfo {author} {\bibfnamefont {J.}~\bibnamefont
  {Cucchiara}}, \bibinfo {author} {\bibfnamefont {S.}~\bibnamefont {Le~Gall}},
  \bibinfo {author} {\bibfnamefont {E.~E.}\ \bibnamefont {Fullerton}}, \bibinfo
  {author} {\bibfnamefont {J.-V.}\ \bibnamefont {Kim}}, \bibinfo {author}
  {\bibfnamefont {D.}~\bibnamefont {Ravelosona}}, \bibinfo {author}
  {\bibfnamefont {Y.}~\bibnamefont {Henry}}, \bibinfo {author} {\bibfnamefont
  {J.~A.}\ \bibnamefont {Katine}}, \bibinfo {author} {\bibfnamefont {A.~D.}\
  \bibnamefont {Kent}}, \bibinfo {author} {\bibfnamefont {D.}~\bibnamefont
  {Bedau}}, \bibinfo {author} {\bibfnamefont {D.}~\bibnamefont {Gopman}},\ and\
  \bibinfo {author} {\bibfnamefont {S.}~\bibnamefont {Mangin}},\ }\bibfield
  {title} {\bibinfo {title} {Domain wall motion in nanopillar spin-valves with
  perpendicular anisotropy driven by spin-transfer torques},\ }\href
  {https://doi.org/10.1103/PhysRevB.86.214429} {\bibfield  {journal} {\bibinfo
  {journal} {Physical Review B}\ }\textbf {\bibinfo {volume} {86}},\ \bibinfo
  {pages} {214429} (\bibinfo {year} {2012})}\BibitemShut {NoStop}%
\bibitem [{\citenamefont {Thiaville}\ \emph {et~al.}(2005)\citenamefont
  {Thiaville}, \citenamefont {Nakatani}, \citenamefont {Miltat},\ and\
  \citenamefont {Suzuki}}]{thiaville_micromagnetic_2005}%
  \BibitemOpen
  \bibfield  {author} {\bibinfo {author} {\bibfnamefont {A.}~\bibnamefont
  {Thiaville}}, \bibinfo {author} {\bibfnamefont {Y.}~\bibnamefont {Nakatani}},
  \bibinfo {author} {\bibfnamefont {J.}~\bibnamefont {Miltat}},\ and\ \bibinfo
  {author} {\bibfnamefont {Y.}~\bibnamefont {Suzuki}},\ }\bibfield  {title}
  {\bibinfo {title} {Micromagnetic understanding of current-driven domain wall
  motion in patterned nanowires},\ }\href
  {https://doi.org/10.1209/epl/i2004-10452-6} {\bibfield  {journal} {\bibinfo
  {journal} {Europhysics Letters}\ }\textbf {\bibinfo {volume} {69}},\ \bibinfo
  {pages} {990} (\bibinfo {year} {2005})}\BibitemShut {NoStop}%
\bibitem [{\citenamefont {Mougin}\ \emph {et~al.}(2007)\citenamefont {Mougin},
  \citenamefont {Cormier}, \citenamefont {Adam}, \citenamefont {Metaxas},\ and\
  \citenamefont {Ferré}}]{mougin_domain_2007}%
  \BibitemOpen
  \bibfield  {author} {\bibinfo {author} {\bibfnamefont {A.}~\bibnamefont
  {Mougin}}, \bibinfo {author} {\bibfnamefont {M.}~\bibnamefont {Cormier}},
  \bibinfo {author} {\bibfnamefont {J.~P.}\ \bibnamefont {Adam}}, \bibinfo
  {author} {\bibfnamefont {P.~J.}\ \bibnamefont {Metaxas}},\ and\ \bibinfo
  {author} {\bibfnamefont {J.}~\bibnamefont {Ferré}},\ }\bibfield  {title}
  {\bibinfo {title} {Domain wall mobility, stability and {Walker} breakdown in
  magnetic nanowires},\ }\href {http://iopscience.iop.org/0295-5075/78/5/57007}
  {\bibfield  {journal} {\bibinfo  {journal} {Europhysics Letters}\ }\textbf
  {\bibinfo {volume} {78}},\ \bibinfo {pages} {57007} (\bibinfo {year}
  {2007})}\BibitemShut {NoStop}%
\bibitem [{\citenamefont {Hubert}\ and\ \citenamefont
  {Schäfer}(2008)}]{hubert_magnetic_2008}%
  \BibitemOpen
  \bibfield  {author} {\bibinfo {author} {\bibfnamefont {A.}~\bibnamefont
  {Hubert}}\ and\ \bibinfo {author} {\bibfnamefont {R.}~\bibnamefont
  {Schäfer}},\ }\href@noop {} {\emph {\bibinfo {title} {Magnetic domains: the
  analysis of magnetic microstructures}}}\ (\bibinfo  {publisher} {Springer
  Science \& Business Media},\ \bibinfo {year} {2008})\BibitemShut {NoStop}%
\end{thebibliography}
%

\end{document}